\documentclass[conf]{new-aiaa}
\usepackage[utf8]{inputenc}
\usepackage{subcaption}
\usepackage{amsmath}
\usepackage{graphicx}
\usepackage[version=4]{mhchem}
\usepackage{siunitx}
\usepackage{bm}
\usepackage{longtable,tabularx}
\usepackage{booktabs}
\usepackage{ulem}
\usepackage{placeins}
\usepackage{comment}

\newif\ifcomments
\commentsfalse
\commentstrue 
\ifcomments
\newcommand{\mmm}[1]{{\color{red}\textbf{MM: #1}}}      
\newcommand{\pjj}[1]{{\color{magenta}\textbf{PJ: #1}}}    
\newcommand{\jz}[1]{{\color{green}\textbf{JZ: #1}}}    
\newcommand{\jn}[1]{{\color{cyan}\textbf{JN: #1}}}    
\else
\newcommand{\mmm}[1]{}
\newcommand{\eag}[1]{}
\newcommand{\pjj}[1]{}
\newcommand{\jz}[1]{}
\newcommand{\jn}[1]{}
\fi

\newcommand{\pd}{\partial} 
\newcommand{\mytens}[1]{\overline{\overline{#1}}}
\newcommand{\myvec}[1]{\boldsymbol{#1}}
\newcommand{\ii}{\mathrm{i}} 

\title{Analysis of a cusped helicon plasma thruster discharge}

\author{Pedro Jim\'enez\footnote{pejimene@ing.uc3m.es \newline}, Jiewei Zhou, Jaume Navarro, Pablo Fajardo, Mario Merino, and Eduardo Ahedo}
\affil{Equipo de Propulsi\'on Espacial y Plasmas, Universidad Carlos III de Madrid, 28911 Legan\'es, Spain}  

\begin{document}

\maketitle

\begin{abstract}
A compact helicon plasma thruster that features a cusp in its internal magnetic field is analyzed with experiments and simulations. 
A compensated Langmuir probe and a Faraday cup are used in the former, while
a hybrid PIC/fluid transport model combined with a frequency-domain electromagnetic field model are used in the latter.
Measurements serve to tune the anomalous transport parameters of the model and overall show the same trends as the numerical results, including a secondary peak of electron temperature downstream in the magnetic nozzle, where electron cyclotron resonance conditions for the 13.56 MHz excitation frequency are met.
The cusp plays a central role in determining the plasma losses to the walls and the profile of electron temperature, which in turn defines the excitation and ionization losses. While losses to the rear wall are reduced, losses to the lateral wall are increased, which, together with the low production efficiency, limit the performance of the device. Shorter chamber lengths and optimization of antenna and cusp location are suggested as potential ways to improve performance.

\end{abstract}

\section{Introduction}

The helicon plasma thruster (HPT) \cite{taka13b,shin14a,trezz13,nava18a} belongs to the family of electrodeless plasma thrusters (EPTs). 
It consists of (1) a cylindrical discharge chamber made of dielectric walls, (2) a gas inlet, usually at the back of the chamber, (3) an external inductor/antenna, emitting the RF power to be absorbed by the plasma, and (4) an external magnetic circuit, generating a magnetic field, which aims to make the plasma transparent to the RF waves, to confine partially the plasma, and to guide an accelerate it externally through a magnetic nozzle (MN) configuration. 
The HPT concept avoids the need for a complementary hollow cathode acting as ion beam neutralizer, the erosion problems linked to electrodes and grids, and simplifies the complex power processing and control unit with respect to mature technologies,
such as Hall Effect and Gridded Ion Thrusters. 
In contrast to these advantages on lifetime and system simplicity, until now, testing on existing HPT prototypes still reports rather low thrust efficiencies\cite{nava18a,siddi17,toson17,taka17b}. 

The conventional magnetic topology of an HPT is quasi-axial inside the chamber \cite{ahed13a,taka19}, thus confining the lateral walls well but permitting large losses on the back wall\cite{ahed13c}. This has led to the investigation of more elaborate topologies, such as those with internal cusps. 
Virko et al. \cite{virk10} studied the impact of different cusped topologies on the acceleration of ions and found that the cusp location relative to the antenna position affects wave accessibility and plasma production.
Similarly, Oshio et al. \cite{oshio18} concluded that positioning the antenna downstream with respect to the null point of the cusp topology slightly increases the thrust performance.

Recently, with the objective of further progressing the HPT technology, a new HPT prototype has been developed and tested with a cusped topology generated by permanent magnets \cite{ruiz22,gomez22}. 
There are two main reasons behind the implementation of permanent magnets instead of solenoids. First, to reduce the power consumption of the thruster system. At this power level, solenoid power consumption would drastically penalize the total efficiency of the system. Second, for the required magnetic field strength, the use of permanent magnets offers a lighter solution with no power expenses. Nevertheless, the magnetic topologies achievable with permanent magnets are typically more complex, e.g. including cusps. We can anticipate that the configuration of the current prototype is far from optimal and the focus of this study is not optimization. Nevertheless, this work provides insightful results inherent to the specific features of this magnetic topology, which are valuable for the further development of permanent magnet-based HPT.
So far, these efforts have followed a dominantly experimental approach, and have lacked a modeling/simulation counterpart to
explain the physics of plasma transport and RF power coupling, which would also shed light on the dominant loss mechanisms and help guide design.

Here, we address these aspects with a numerical and experimental investigation of the cusped HPT prototype of \cite{ruiz22}. 
Section \ref{sec:exp_setup} of the paper presents a set of vacuum chamber measurements of the plasma density, temperature, and electric potential obtained in the plasma plume. The posterior sections apply an axisymmetric simulation code to fit those measurements and then investigate the plasma discharge properties inside the chamber and the plume, and the related performances. Special attention is given to the current and power balances, in order to quantify the different sources of mass and energy losses. 

The simulation code used is HYPHEN-EPT, which includes different modules to model the plasma-wave energy deposition process [W(ave)-module], and the production and quasi-steady transport of the several plasma species
[I(on)- and E(lectron)-modules]. The I-module operates on a PIC formulation of several heavy species (ions and neutrals), and the E-module deals with a diffusive, magnetized fluid model for electrons. These modules are common for simulations of EPTs and Hall-effect thrusters and have been described in detail in recent publications \cite{zhou19a,zhou22a,pera22b}, and briefly overviewed in Section \ref{sec:tpt}. The W-module deals with the Maxwell equations for the RF electromagnetic fields, assuming a quasi-steady cold plasma dielectric tensor, determined from the I- and E-modules. Reciprocally, the W-module provides the energy-deposition map for the E-module. All modules run sequentially to achieve a self-consistent steady-state solution.

The W-module builds on and improves previous models. Tian et al. \cite{tian18a} solved the Maxwell equations in the frequency domain with a finite difference (FD) model in the source and near plume of an HPT with quasi-parallel magnetic topology. Jiménez et al. \cite{jime22a} extended the FD model to include RF propagation in the far plume and found a significant power deposition per particle at the downstream electron-cyclotron resonance surface,
a fact corroborated recently by the experimental measurements of Vinci et al. \cite{vinci23a}. 
Sánchez-Villar et al. \cite{svil21a} proposed an alternative finite element (FE) model for the W-module, although limited to the axisymmetric azimuthal mode number $m=0$.  This was applied to simulate
a coaxial electron cyclotron resonance thruster (ECRT). A combined experimental-numerical analysis of the ECRT, similar to the one proposed here for the HPT, followed in \cite{svil23}. 
Despite the increased complexity, FE formulations are more suited to general geometries and computationally more efficient than FD formulations. 
As part of the contributions of the present work, in Section \ref{sec:wmodule}, the FE model is extended to any azimuthal mode number $m$, as required for application to an HPT with a helical antenna (with $m= 1$ as the dominant azimuthal wave numbers).  

The electron formulation within the E-module makes use of some empirical parameters to adjust the anomalous transport properties, known from experiments but still elusive to consistent fluid modeling.  
Section \ref{sec:fitting} compares
the experimental and numerical results for the best fit of the above empirical parameters. 
Section \ref{sec:sensitivity} presents a sensitivity analysis of the numerical results to those parameters. Sections \ref{sec:wave} to \ref{sec:perf} discuss the EM wave fields, the power deposition, the maps of plasma magnitudes, and the thruster performances, identifying the main causes of
low efficiency. Section \ref{sec:concl} summarizes the conclusions.
Appendix \ref{sec:appendix} compares the latest FE and FD formulations of the W-module.
  
\section{Device and experimental results}\label{sec:exp_setup}

\begin{figure}[ht]
    \centering
    \includegraphics[width=0.5\textwidth]{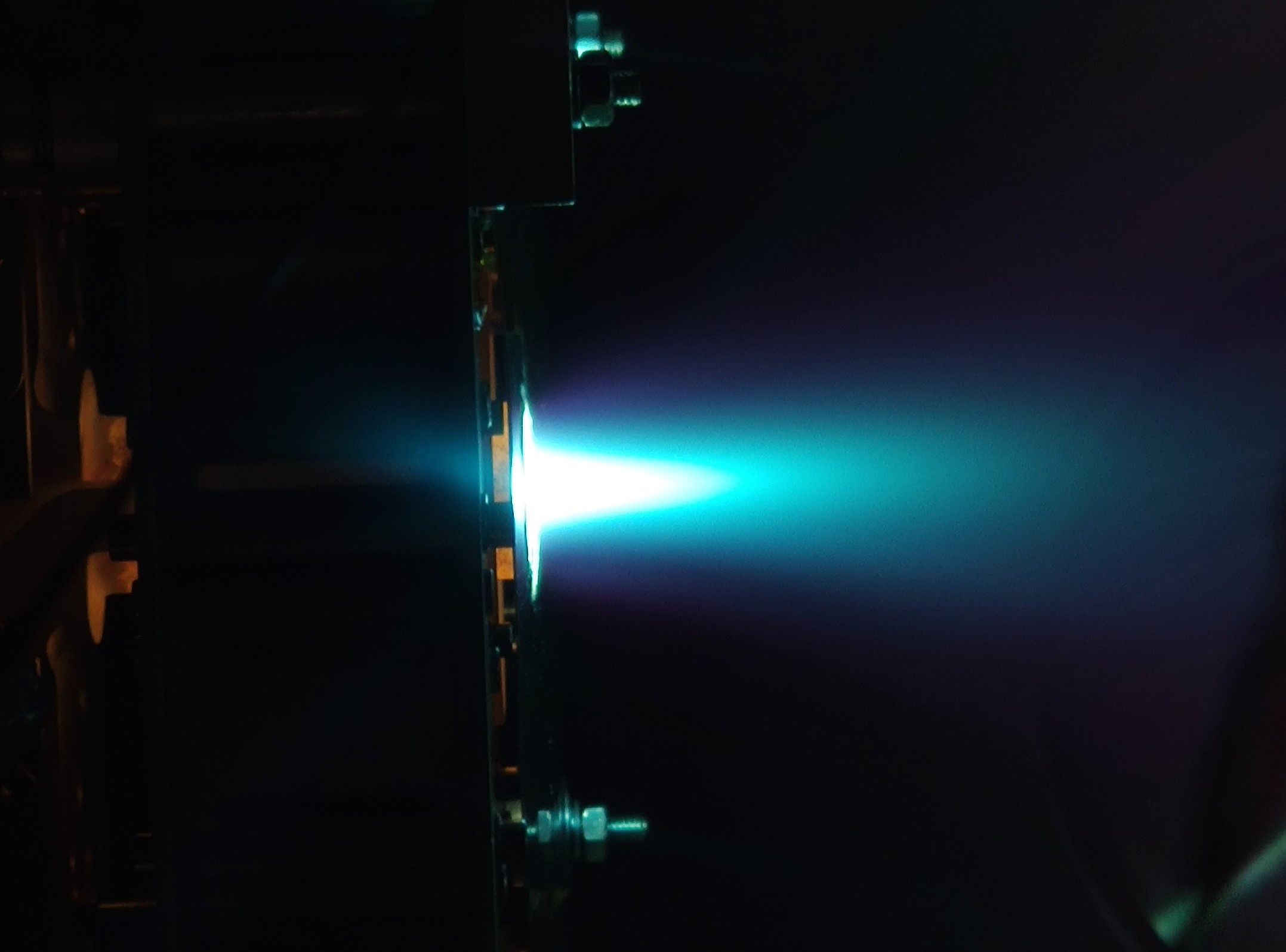}
    \caption{HPT prototype lateral view, operating with Xe in the UC3M lab vacuum chamber.}
    \label{fig:thruster}
\end{figure}

The device under study is the thruster unit breadboard model developed jointly by Sener Aeroepacial and Universidad Carlos III de Madrid \cite{ruiz22}. The device is a compact 400W-class HPT with a 12.5 mm source radius and a 60mm source length, made of boron nitride. Xenon is injected through a multi-hole injector plate, which is embedded at the back side of the discharge tube. RF power is fed to a half-helical antenna wrapped around the source. RF power is generated using an industrial power supply, an RF Seren 2001 amplifier, and is coupled to the antenna through a customized matching network in order to match the system impedance for all operating conditions, with less than 2\% of reflected power. The device under vacuum chamber operation is shown in figure \ref{fig:thruster}, and its design and operational parameters are shown in the first part of table \ref{table:simparams}.

\begin{figure}[ht!]
\centering
\begin{minipage}[c]{0.47\textwidth}
\includegraphics[width=\textwidth]{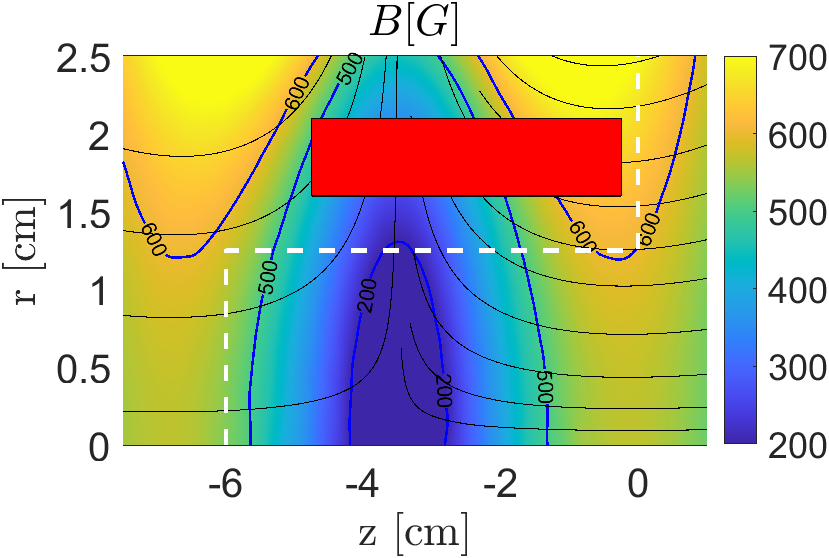}
\end{minipage}
\begin{minipage}[c]{0.5\textwidth}
\includegraphics[width=\textwidth]{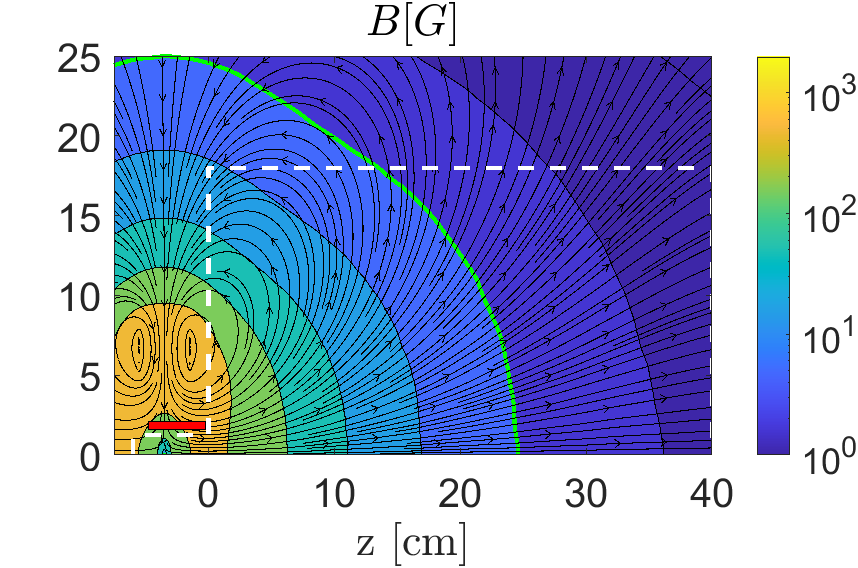}
\end{minipage}
\caption{Applied Magnetic Field (azimuthally-averaged). Zoom in to the source region (left) and full wave domain (right). The antenna is depicted in red and the boundary of the transport domain is represented by a white dashed line. The green line is the Electron Cyclotron Resonance (ECR) surface.
}
\label{fig:Ba}
\end{figure}

The magnetic field is generated by a set of radially-polarized neodymium magnets, which are assembled in an annular aluminum support. The width of the magnet assembly is 40 mm and the inner radius is 50 mm. This magnet arrangement generates a cusp with a separatrix plane inside the source, $35$ mm upstream of the outlet. The azimuthally-averaged magnetic field is shown in Figure \ref{fig:Ba}. The forward peak of the on-axis magnetic field occurs at the
magnetic nozzle (MN) throat, coinciding with the source outlet section, and is about 750 G. The rear peak of the on-axis magnetic field lies behind the backplate injector.
The rationale for this setup was to avoid the use of bulky and power-consuming solenoids, which were used in previous developments \cite{nava18a}.


\subsection{Setup description}

\begin{figure}[ht!]
    \begin{minipage}[c]{0.49\textwidth}
    \includegraphics[width=1\textwidth]{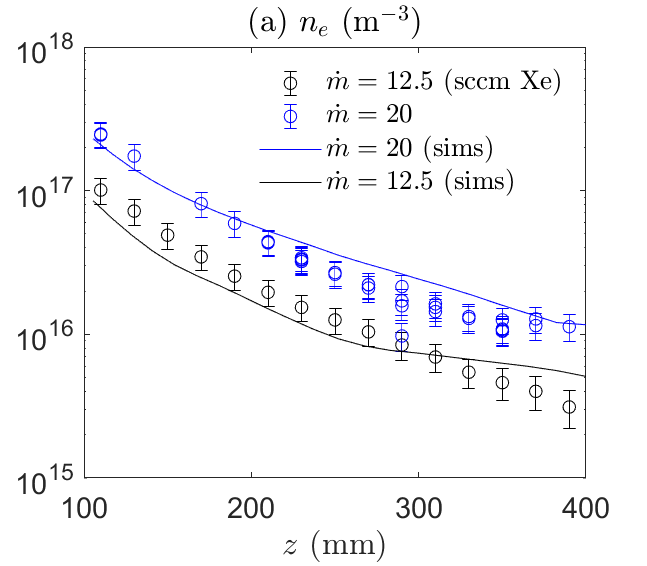}
    \vspace{0cm}\hspace{2 cm}
    \end{minipage}
    \begin{minipage}[c]{0.49\textwidth}
    \includegraphics[width=1\textwidth]{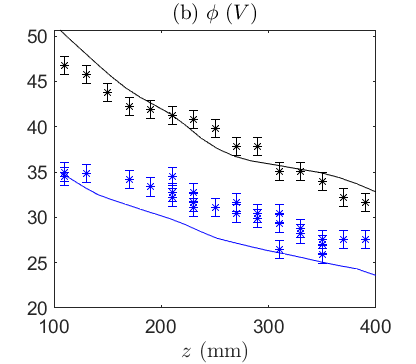}
    \vspace{0cm}\hspace{2 cm}
    \end{minipage}
    \begin{minipage}[c]{0.49\textwidth}
    \includegraphics[width=1\textwidth]{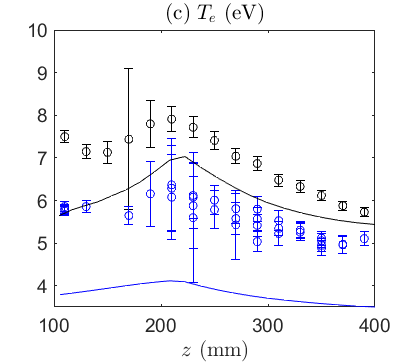}
    \vspace{0cm}\hspace{2 cm}
    \end{minipage}
    \begin{minipage}[c]{0.49\textwidth}
    \includegraphics[width=1\textwidth]{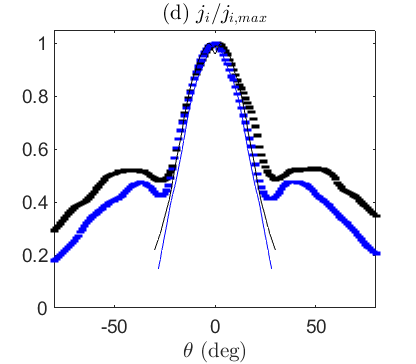}
    \vspace{0cm}\hspace{2 cm}
    \end{minipage}
    \caption{Experimental (symbols) and numerical profiles (solid lines) for electron density $n_e$ (a), electrostatic potential $\phi$ (b), electron temperature $T_e$ (c), and ion current density $j_i$ (d).
    Results are for two different mass flows $\dot{m}=12.5$ sccm (black) and $\dot{m}=20$ sccm (blue), at constant RF power 450 W. Error bars consider only the propagation of instrument uncertainty, except for the density where $20\%$ minimum uncertainty is assumed.
    }\label{fig:experiments}
\end{figure}

\setlength{\arrayrulewidth}{0.8pt}
\begin{table}[ht!]
\centering
\begin{tabular}{ccc}
\toprule[2pt]
Parameter & Symbol  & Value \\
\midrule[1.5pt]
\setlength{\arrayrulewidth}{0.4pt}
Thruster length & $L$  & $6$ cm\\
Thruster radius & $R$ & $1.25$ cm\\
Injector radius & $R_{inj}$ & $0.625$ cm \\   
Mass flow rate & $\dot m$ & $12.5$--$20$ sccm \\   
Antenna frequency        & $f = \omega/(2\pi)$    & 13.56 MHz             \\
Antenna Power            &          $P_a$         & 450 W           \\
Antenna loop radius      & $r_a$                & 1.85 cm             \\
Antenna length           & $l_a$                &  4.5 cm               \\
Antenna central position & $z_a$           & -2.5 cm             \\
Antenna thickness        & $d_t$                & 0.5 cm              \\
\midrule
Plume length & $L_p$ &40 cm\\
Maximum plume radius & $R_p$ & 18 cm \\
I-mesh size & - & 4961 cells \\
E-mesh size & -  & 3671 cells\\
I-module time step & $\Delta t_I$  & $2.5\cdot10^{-8} s$ \\
E-module time step &  $\Delta t_E$  & $5\cdot10^{-9} s$ \\
W-module update time-step  &       $\Delta t_W$    &   $1.25\cdot10^{-6} s$ \\
Total simulation time             & $t_{sim}$ & 3.75 ms \\

Rectangular wave mesh    &          -           &  $z\in[-13.5, 43]$cm, $r\in[0,25]$ cm\\
W-mesh size              &          -             & $10^{6}$ cells         \\
\midrule
Maximum Applied Magnetic Field & $B_a$  & $10^{3}$  G\\
Gyro-frequency   &  $\omega_{ce}$ &  $10^7$-$10^{10}$ $\mathrm{s}^{-1}$\\ 
Density & $n_e$  &   $10^{15}$-$10^{20}$ $1/\mathrm{m}^3$\\
Debye Length & $\lambda_D$ & $10^{-6}$-$10^{-3}$ m\\ 
Temperature & $T_e$    & 2-6 eV \\
Effective Collision Frequency & $\nu_e$  &  $10^{4}$-$10^{8}$ $\mathrm{s}^{-1}$\\

\bottomrule[2pt]
\end{tabular}
\caption{Cusped HPT parameters (upper part of the table), numerical parameters used in the simulation (mid part of the table) and characteristic plasma conditions in the simulation domain (bottom part of the table).
\label{table:simparams}
}
\end{table} 

The experimental results presented next are part of the first coupling test campaign of the propulsion system assembly, which, in addition to the thruster unit, includes a dedicated power processing unit and a gas valve optimized for this thruster. 
The thruster prototype has been tested in a $1.5$ m diameter, $3.5$ m length stainless steel vacuum chamber. The ultimate dry vacuum is $10^{-7}$ mbar, while the background pressure stands at  $5.9\cdot10^{-6}$ to $4.6\cdot10^{-5}$ mbar for a flow rate of Xe between 5 and 50 sccm.
Additional results of interest for this work are the Laser induced fluorescence (LIF) measurements of the plasma plume, recently presented in \cite{vinci22c}.

Two operating points have been tested. RF power has been kept constant at 450 W, while the Xe flow rate has been set at 12.5 sccm (low flow rate case) and 20 sccm (high flow rate case).
A set of electrostatic probes was mounted on a radial-polar robotic arm, which allows us to inspect intrusively the plasma properties within a semicircular horizontal plane, $\rho \in (0,400)$ mm, $\theta \in (-\pi/2, \pi/2)$, centered at the thruster outlet. The set of probes includes a radio-frequency compensated Langmuir probe \cite{sudi94} to infer plasma density, electron temperature, and plasma potential along the axis line, 100-400 mm downstream from the thruster outlet. The second probe is a Faraday cup that has been used to characterize the ion current density along a semi-arch of radius $\rho=350\;\si{mm}$ \cite{wijnen22}. 

\subsection{Experimental results}

The main results are shown in
Figure \ref{fig:experiments}, together with the anticipated numerical
results.
We discuss here the experimental profiles and leave for
Section \ref{sec:fitting} the comparison with the simulation results. 
For all experimental profiles, error bars were estimated by error propagation of the instrument uncertainties and the parameters involved in the measurement of each physical property. Importantly, for LP measurements, the error bars do not include the uncertainty on the theoretical model as in \cite{lobb17}. For the specific case of the ion density, strongly affected by the uncertainty in the effective collection area, a minimum uncertainty of 20 \% is assumed. For the rest of the properties, no additional uncertainty has been added.

The electron density profile on the axis is shown in Figure \ref{fig:experiments}(a). The electron density drops almost 2 orders of magnitude in the explored range. It is seen to decrease at a slightly lower rate in the high mass flow case. 
Plasma potential and electron temperature are shown in panels (b) and (c) of Figure \ref{fig:experiments}, respectively. The trend of potential is monotonically decreasing, as expected in a MN \cite{meri16a}. The electron temperature is larger in the low-mass flow rate case, and consistently, the potential drop is also larger in that case. 
The electron temperature profile shows a mild peak at 200--250 mm of the source, contrasting with the typically monotonically decreasing profiles reported in the literature \cite{litt19, mart15a}.

Finally, the normalized profiles of the ion current density are shown in panel (d). The current density is measured with the Faraday probe in a semi-arch of radius $\rho=350\;\si{mm}$. The ion beam features a single symmetric peak with lateral wings. The first indicates the formation of a well-collimated ion beam, while the second could be the consequence of the interaction of the plasma with the background pressure, or the presence of secondary beams linked to heating/ionization in the periphery of the main plasma column. 

\section{Simulation model}\label{sec:mod}

The HYPHEN-EPT simulation model is composed of different modules to solve the plasma profiles and electromagnetic fields in the meridian plane. 
First, there is the E-module that solves for the slow dynamics of electrons; a magnetized, drift-diffusive, quasineutral fluid model \cite{ahed23a,zhou22a}, with finite difference and finite volume techniques.
Second, there is the I-module that solves for the dynamics of
ions and neutrals, using a PIC formulation and a Monte Carlo collision scheme \cite{domi18c,domi21a}. The I-module determines the quasinuetral electron density used in the E-module. The E- and I-modules are completed by a S(heath)-module for the nonneutral Debye sheaths around all the walls of the integration domain.
Together, the E-, I- and S-modules constitute the slow plasma transport part of HYPHEN-EPT. 
In the present study, we consider the quasineutral, low-$\beta$ limit of the plasma transport problem.
Quasineutrality is a key element of the transport model to keep the computational cost affordable in a workstation. The low$\beta$ assumption allows neglecting the stationary plasma-induced magnetic field with respect to the applied one.

\begin{figure}[ht]
    \centering
    \includegraphics[width=0.8\textwidth]{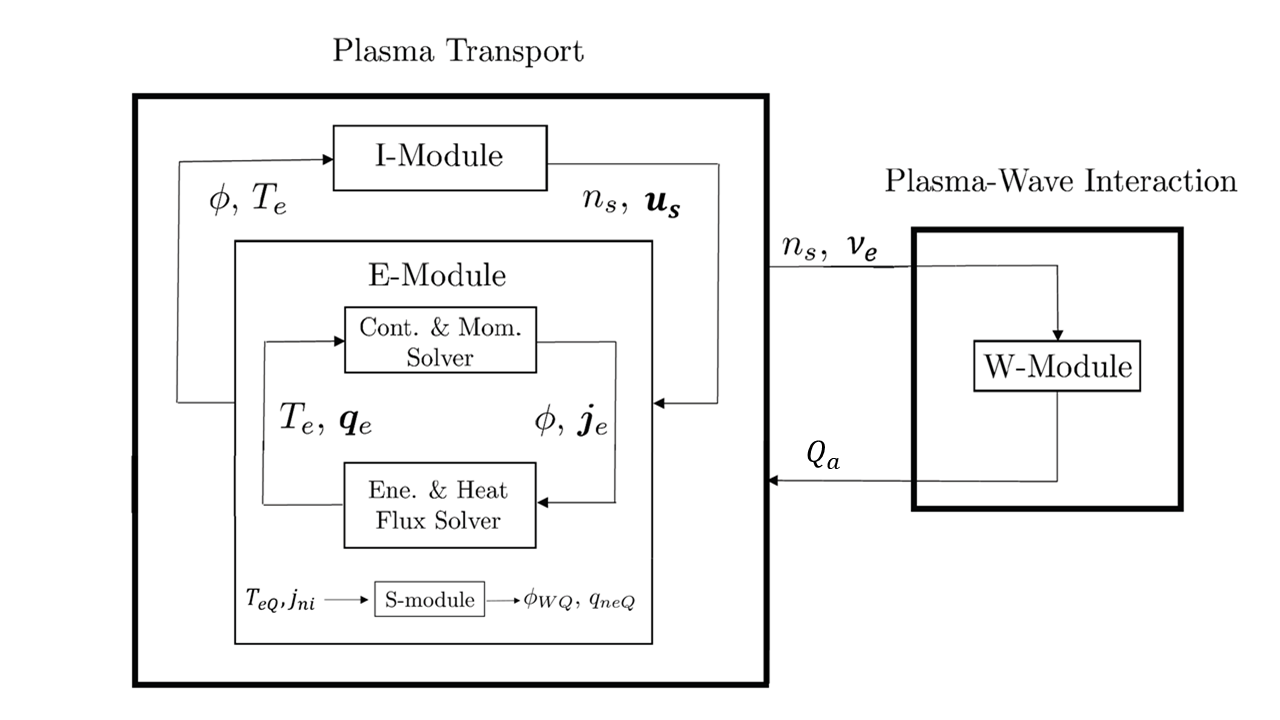}
    \caption{Simulation model structure, modules and interfaces.}
    \label{fig:codeSturct}
\end{figure}

Third, the W-module solves Maxwell's equations, with finite element discretization in the frequency domain, for the high-frequency electron-wave interaction and the subsequent deposition of wave energy into the electrons.
Figure \ref{fig:codeSturct} shows a sketch of main model blocks and the plasma magnitudes (defined later in this Section) acting as inputs/outputs between them.

\begin{figure}
    \centering
    \includegraphics[width=0.8\textwidth]{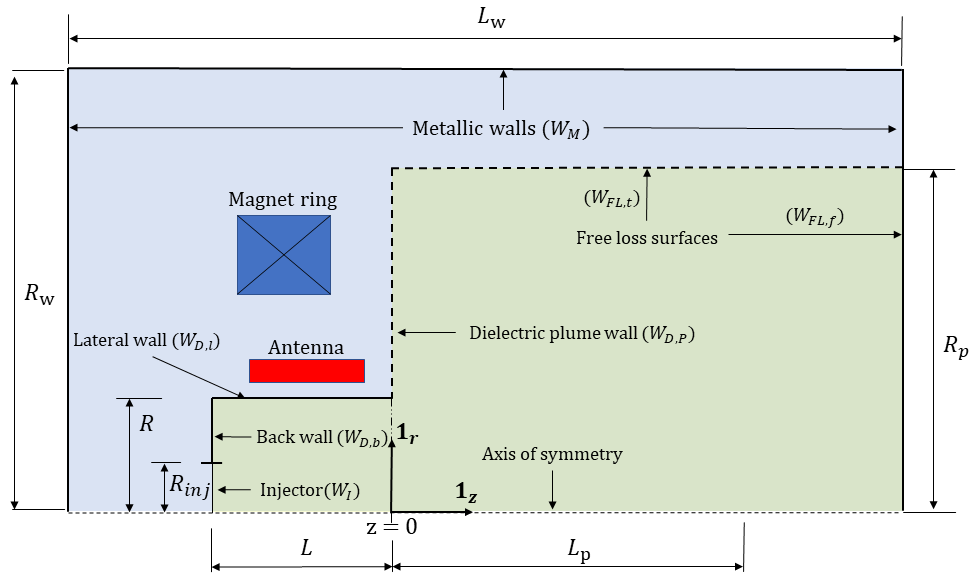}
    \caption{Simulation domain and boundary conditions (not to scale). The transport domain corresponds to the green area. The W-module domain is the union of the green and blue areas.
    }
    \label{fig:sim_setup} 
\end{figure}

The second part of table \ref{table:simparams} summarizes the simulation parameters. Two simulations, corresponding to the cases with mass flow rates
$12.5$ and $20$ sccm are run.
An overview of the computational domain is shown in figure \ref{fig:sim_setup}.
The I-module 
uses a nonuniform Cartesian-type mesh, 
adapted to the expected plasma gradients and to keep a statistically acceptable number of macroparticles per cell.
The E-module integrates the fluid equations in
a magnetic field aligned mesh (MFAM) to minimize numerical diffusion in the anisotropic slow dynamics of the magnetized electrons
\cite{pere16,zhou19a}. Finally, the W-module uses a much finer unstructured mesh, tailored to resolve the small wavelengths present in the problem.
Linear interpolation between the meshes is used to communicate the various modules.

As in \cite{jime22a}, the simulation domain
includes the electron-cyclotron resonance (ECR) surface present in the MN
(the green line in Figure \ref{fig:Ba}),
located where the cyclotron frequency of the applied magnetic field is 13.56 MHz.

The numerical model is run sequentially until steady state is reached,
both in thrust and other performance figures; this takes place after 3.75 ms of physical time. 
As in \cite{zhou22a}, the simulation is pre-initialized by calling the I-module alone, with a Boltzmann relation-like model for the electrostatic potential. This allows the simulation domain to be quickly filled with macroparticles that come from the injector before calling the E- and W-modules. 
The main timestep $\Delta t_I$ is the one advancing the I-module. The E-module generally requires 5-10 subiterations, each with timestep $\Delta t_E$. In contrast, the W-module, which solves the fields in the frequency domain, is called after a fixed number of transport timesteps, about each 50-250 depending on the problem, in order to update the cycle-averaged power deposition. Customary to PIC simulations, the results presented here are time-averaged over 200 I-module steps to filter noise and short period oscillations.

In the following, additional details of the transport and wave parts of HYPHEN-EPT are provided. The cylindrical vector basis $\{\bm1_z, \bm1_r, \bm1_\theta\}$ and the magnetic vector basis
$\{\bm1_\parallel,\bm 1_\perp,\bm 1_\theta\}$, with 
$\bm1_\parallel=\myvec{B}/B$ and
$\bm 1_\perp =\bm1_\theta \times \bm1_\parallel$, are used.
  
\subsection{Plasma transport model}\label{sec:tpt}
 
The reader is referred to \cite{zhou22a} for a full description of the hybrid transport  model used in this work; in the following, only an overview of the key characteristics for this work is given.

The E-module integrates the following fluid equations for electrons:
\begin{equation}
\label{eq:quasy}
    n_{e}=\sum_{s \neq e} Z_{s} n_{s},
\end{equation}
\begin{equation}
\label{eq:cont}
    \nabla \cdot \boldsymbol{j}_{e}=-\nabla \cdot \boldsymbol{j}_{i},
\end{equation}
\begin{equation}
\label{eq:moment}
    0=-\nabla\left(n_{e} T_{e}\right)+e n_{e} \nabla \phi+\boldsymbol{j}_{e} \times \boldsymbol{B}+\boldsymbol{F}_{r e s}+\boldsymbol{F}_{ano},
\end{equation}
\begin{equation}
\label{eq:energy}
    \frac{\partial}{\partial t}\left(\frac{3}{2} n_{e} T_{e}\right)+\nabla \cdot\left(\frac{5}{2} T_{e} n_{e} \boldsymbol{u}_{e}+\boldsymbol{q}_{e}\right)=-\nabla \phi \cdot \boldsymbol{j}_{e}-Q_{inel}+Q_a,
\end{equation}
\begin{equation}
\label{eq:heat}
    0=-\frac{5 n_{e} T_{e}}{2 e} \nabla T_{e}-\boldsymbol{q}_{e} \times \boldsymbol{B}+\boldsymbol{Y}_{\text {res }}+\boldsymbol{Y}_{\text {ano }},
\end{equation}
where the unknowns are $n_e$, $\phi$, $\boldsymbol{j_e}$, $T_e$ and $\boldsymbol{q_e}$.
The I-module provides the right hand side of Eqs.(\ref{eq:quasy},\ref{eq:cont}). 
In the energy equation, $Q_{inel}$ 
is the power density lost by electrons during inelastic collisions
(i.e. excitation and ionization) and
$Q_a$ is the volumetric power density deposited by the electromagnetic fields, which is furnished by the W-module.
 
In the momentum and heat flux equations, the resistive terms are
\begin{equation}
    \boldsymbol{F}_{r e s}=- m_e n_e 
    \Big(\nu_e  \boldsymbol{u}_e- \sum_{s \neq e}\nu_{e s} \boldsymbol{u}_{s}\Big),
       \qquad 
    \boldsymbol{Y}_{r e s}=-
   \frac{m_{e} \nu_{e}}{e} \boldsymbol{q}_{e}
\end{equation}
where $\nu_{e s}(T_e)$ is the collision frequency with heavy species $s$, and $\nu_{e}=\sum_{s \neq e} \nu_{e s}$ is the total collision frequency of electrons. 
The terms representing anomalous transport are modeled here as
\cite{zhou22a}
\begin{equation}
    \boldsymbol{F}_{ano}=\alpha_{ano} B j_{\theta e} \mathbf{1}_{\theta},
    \qquad
    \boldsymbol{Y}_{\text {ano }}=-\alpha_{\text {ano }} B q_{\theta e} \mathbf{1}_{\theta}-\left(m_{e} \nu_{q} / e\right) q_{\| e} \mathbf{1}_{\|},
\end{equation}
where both $\alpha_{ano}$ and $\nu_{q}$ are phenomenological parameters that will be fitted later with the experimental results. The terms with $\alpha_{ano}$ represent the effect of high-frequency turbulence on quasi-steady electron transport, while the term with $\nu_{q}$ empirically reduces heat fluxes in the parallel direction, to capture the collisionless cooling observed in magnetically expanded plasma plumes \cite{litt16,ahed20a}. 
Observe that the drift-diffusive equations (\ref{eq:moment}) and (\ref{eq:heat}) provide generalized Ohm and Fourier laws
for $\boldsymbol{j}_{e}$ and $\boldsymbol{q}_{e}$. 
The numerical integration of the electron model in the MFAM is based on a semi-implicit scheme for time discretization and finite volume/gradient reconstruction methods for spatial discretization \cite{zhou19a, zhou22a}. 

The S-module allows matching the boundaries of the quasi-neutral domain to either the thruster walls
(using a sheath model)
or the downstream plasma plume
(using matching conditions with infinity) \cite{domi22a}.
Provided the electron temperature $T_e$ and the current densities of the ion species at the edge of the sheath, the S-module
fixes the necessary electron current and heat fluxes and the local potential fall $\phi_{WQ}$ between a quasineutral boundary point Q and a wall point W (or infinity). 
The S-module accounts for
material type, recombination, and secondary electron emission by providing different plasma-wall interaction models and fitting parameters. 
For the ceramic walls of the thruster, we implement locally $\boldsymbol{j_e} + \boldsymbol{j_i}$ = 0.
For downstream free loss (FL) surfaces,  
we determine a potential at infinity $\phi_\infty$
such that the plume is current-free globally, i.e. $\iint_{FL}\boldsymbol{j}\cdot\boldsymbol{n} \mathrm{d}S = 0$,
with the integration domain extended to the whole free-loss (FL) surface
\cite{domi22a}.

In this work, Xe, Xe$^+$, and Xe$^{++}$ species are included 
in the I-module. Xenon is introduced through the injector surface on the back wall of the thruster, shown in figure \ref{fig:sim_setup}. 
The number of particles injected per time step is selected and population control strategies \cite{domi18c} are enforced to achieve a statistically sufficient number of particles per cell in the periphery of the plume. Not less than 100 particles are observed in each PIC cell in steady state.
The number of total particles of all ion and neutral species in steady state approaches $4\cdot10^6$.  

Finally, the collision mechanisms considered in this work are: ionization and excitation of singly and doubly charged ions, elastic electron-neutral and electron-ion collisions, and wall recombination of all ion populations.
 
\subsection{Electromagnetic wave model}\label{sec:wmodule}

The full-wave, frequency domain, cold-plasma model presented
in \cite{jime22a} is used here for the W-module. 
Rather than a finite difference discretization as in that reference, we adopt a discretization of mixed Lagrange / vector finite elements similar to that in \cite{svil21a}. 
The main advantages of this change of approach are discussed in the Appendix \ref{sec:appendix}.
A major feature of the new implementation is that the present code can simulate arbitrary azimuthal modes $m$, while that of \cite{svil21a} only handles $m=0$. This is central for HPT simulation, where the $m=1$ mode is prevalent.
  
The wave equation for the fast electric field $\bm{\mathcal E}$,
\begin{equation}\label{eq:wave}
\varepsilon_0\mu_0\frac{\pd^2 \bm{\mathcal{E}}}{\pd t} + \nabla \times(\nabla \times \bm{\mathcal{E}})= -\mu_0(\mathcal{\bm J}_p + \mathcal{\bm J}_a  ),
\end{equation}
is solved for harmonic solutions of the form
\begin{equation}
\bm{\mathcal{E}}(z,r,\theta,t) = \Re \left[\bm{E} (z,r)\exp(-\ii \omega t + im\theta)\right].
\end{equation}
Similar expansions are used in $\mathcal{\bm J}_p$ and $\mathcal{\bm J}_a$,
the plasma and antenna current densities.
Ignoring the slow ion response to the wave fields, the complex magnitude of the plasma current, $\bm J_p$, is expressed 
using the cold plasma dielectric tensor formalism 
\cite{STIX92}  as:
\begin{align}
\bm J_p = -i\omega\varepsilon_0(\mytens\kappa - \mytens 1) \cdot \bm E,
\end{align}
where $\mytens 1$ is the identity tensor and
the $\mytens\kappa$ the dielectric tensor, which takes the following form in the $\{\bm1_\parallel,\bm 1_\perp,\bm 1_\theta\}$ vector basis:
\begin{equation}
\mytens{\kappa}(z,r)=
\begin{pmatrix} 
P & 0 & 0 \\ 
0 & (R+L)/2 & -i(R-L)/2 \\ 
0 & i(R-L)/2 & (R+L)/2 
\end{pmatrix}, 
\label{eq:kappa}
\end{equation}
with
\begin{equation*}  
R=  1-\frac{\omega_{pe}^2}{\omega(\omega+\ii\nu_e-\omega_{ce})},
\qquad
L=  1-\frac{\omega_{pe}^2}{\omega(\omega+\ii\nu_e+\omega_{ce})},
\qquad
P= 1-\frac{\omega_{pe}^2}{\omega(\omega+\ii\nu_e)}.
\end{equation*} 
The electron cyclotron and plasma frequencies, 
$\omega_{ce}= e B_a/ m_e$ and $\omega_{pe}= \sqrt{ne^2/(m_e\varepsilon_0})$, are the main plasma parameters in the electromagnetic model, 
while the electron collisionality $\nu_e$ is an input from the slow plasma transport code. 

The weak form of equation \eqref{eq:wave} after expanding in $\omega$ and $m$ is:
\begin{equation}
\begin{aligned}
&\iint_{\Omega}\left\{\left[\left(\nabla_{t}+\myvec{1_\theta} \frac{i m}{r}\right) \times \mathbf{T}^{(m)}\right] \cdot \left[\left(\nabla_{t}-\myvec{1_\theta} \frac{i m}{r}\right) \times \mathbf{E}^{(m)}\right]-k_{0}^{2} \mathbf{T}^{(m)} \cdot \mytens{\kappa} \cdot \mathbf{E}^{(m)}\right\} dS +\\
&+i k_{0} \int_{\delta\Omega}\left[\hat{n} \times \mathbf{T}^{(m)}\right] \cdot\left[\hat{n} \times \mathbf{E}^{(m)}\right] dl =i\mu_0\omega\iint_{\Omega} \mathbf{T}^{(m)} \cdot \myvec{J_a}^{(m)}ds
\end{aligned}
\end{equation}
where $\Omega$ and $\delta\Omega$ are the simulation domain and its enclosing boundary, and $\myvec{T}^{(m)}$ is a test function. As in \cite{svil21a}, a mixed element discretization is used with a Nédélec basis for $(E_z,E_r)$ and a Lagrange basis for $E_\theta$.
This conforming mixed element approach has proven to be successful in preventing artificial accumulation of spurious current \cite{nede80,MONK03}.

The time-averaged power density deposited into the plasma, which is the main input to the transport code from the W-module, is
\begin{align}
Q_a^{(m)} = \frac{1}{2}\mathfrak{R}\left((\bm{J}_{p}^{(m)})^* \cdot \bm{E^{(m)}}\right).
\end{align}
for each azimuthal mode $m$.
Due to orthogonality, each mode can be solved separately. Previous studies on HPTs using helical antennas have shown the prevalence of $m=1$ in total plasma resistivity \cite{tian17a, zhou19a, chen03}. This prevalence is also confirmed for the current cusped configuration later in this work, and therefore only the $m=1$ mode will be considered in the results.

 
Finally, to close the problem, the antenna current density, $\bm J_a$, and the boundary conditions must be given as input to the W-module.
A half-helical antenna, identical to the one reported in \cite{jime21a}, is used. 
The W-module domain, shown in figure \ref{fig:sim_setup} is larger than the transport domain, adding a vacuum region around it, and terminated at perfect electric conductor (PEC) walls (which would mimic a small vacuum chamber). On the axis of symmetry, regularity/smoothness boundary conditions are used \cite{jime22a}. 

The selection of the call frequency of the W-module (i.e., the number of PIC steps per wave call) is a compromise between very frequent EM calls and numerical stability.
The local mesh size for the W-module is selected so that there are at least 20 nodes per wavelength in average.
This allows one to resolve the fine wave structures \cite{jime22a}.
Additionally, to avoid noise at sharp transitions in plasma density \cite{jime21a}, the effective electron collisionality has been smoothly reduced from its computed value to 0 in the last $0.05R$ by the lateral and back walls of the thruster. The deposited power profile has been checked to be insensitive to small variations in the thickness of the numerical transition region around this value. 
  
\section{Comparison of experimental and numerical results}
\label{sec:experimental}
 
In Section \ref{sec:fitting}, the experimental measurements are used to determine the value of the empirical parameters $\alpha_{ano}$ and $\nu_q$ in the transport code, while a sensitivity analysis for these two parameters is presented in section \ref{sec:sensitivity}.
 
\subsection{Model fitting}\label{sec:fitting}

The experimental data has been used
to fit
$\alpha_{ano} = 0.0165$ and $\nu_q=3.5\cdot10^{7}\;\si{s^{-1}}$ in the simulations. 
The agreement on plasma electron density $n_e$ can be discussed
in view of figure \ref{fig:experiments}(a).
The expansion rate agrees well in the first part of the expansion $\dot m=12.5$ sscm, while it is marginally slower than the experiments for $\dot m= 20$ sccm.

Once adjusted to a common reference, the electrostatic potential $\phi$ in Figure \ref{fig:experiments}(b) shows fairly good agreement between the experiments and simulations.
  
In both $n_e$ and $\phi$, a slight slope change is observed downstream in the simulated results for $\dot m=12.5$ sscm that is not recorded in the experimental measurements. 
This could be related to a minor decrease in the divergence rate of the ion streamlines observed in the last part of the domain (not shown).
  
The electron temperature $T_e$ is shown in Figure \ref{fig:experiments}(c). Although there is a bias between experiments and simulations, the trends and the presence of the maximum are similar in both. 
This local maximum coincides well with the location of the ECR (see Figure \ref{fig:Ba}), which is a secondary location for the absorption of RF power by the electrons.
The cooling rate after the maximum is well captured by the model.
However, the curve for $T_e$ partially misses the behavior of the experimental data in the vicinity of the source exit, especially at $\dot m=12.5$ sccm.

Finally, the normalized ion current density, including singly and doubly charged contributions, in Figure\ref{fig:experiments} (d) shows good agreement between experimental and numerical results at low angles from the centerline, up to $\pm 25$ deg. The lateral wings at high angles, which are present in the experiments, are not captured by the simulations. 
The ion current and plasma density in these regions is expected to depend on the background pressure of the vacuum chamber \cite{wach20b}, which is not modeled in the simulations.
The ion current profile in the simulation is also known to be highly sensitive to anomalous cross-field diffusion, controlled here by $\alpha_{ano}$, and to conditions upstream at the source \cite{zhou22a}.



\subsection{Sensitivity analysis} \label{sec:sensitivity}

To assess the sensitivity of the simulations to changes in phenomenological parameters $\alpha_{ano}$ and $\nu_q$, a parametric study is performed for the case $\dot{m} = 12.5$ sscm.
According to \cite{zhou22a}, 
the anomalous transport parameter $\alpha_{ano}$ 
affects the cross-field transport, especially in near-collisionless regions such as the plume. Increasing $\alpha_{ano}$ increases plume divergence and plasma losses to the walls inside the thruster.
Increasing the anomalous collisionality $\nu_q$ in the heat flux equation reduces the parallel conductivity, making magnetic lines less isothermal. A higher $\nu_q$ increases the electron cooling rate in the plume and also enables a higher $T_e$ inside the thruster.

\begin{figure}[ht!]
\centering
    \begin{minipage}[c]{0.49\textwidth}
    \centering
    \includegraphics[width=0.95\textwidth]{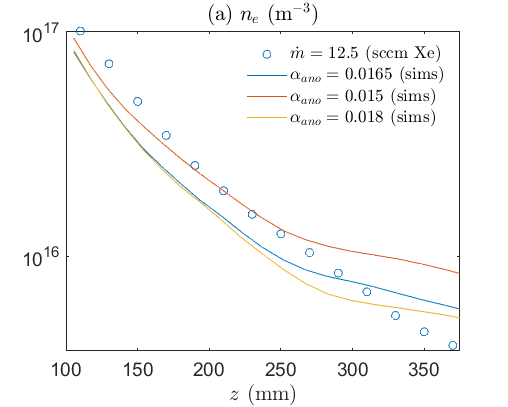}
    \vspace{0cm}\hspace{2 cm}
    \end{minipage}
    \begin{minipage}[c]{0.49\textwidth}
    \centering
    \includegraphics[width=0.95\textwidth]{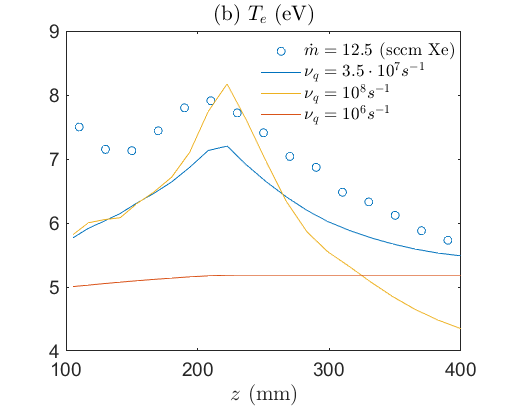}
    \vspace{0cm}\hspace{2 cm}
    \end{minipage}
    \caption{Sensitivity analysis on the plasma density and electron temperature with varying $\alpha_{ano}$ and $\nu_q$.
    }\label{fig:sensitivity}
\end{figure}

Figure \ref{fig:sensitivity}(a) shows the plasma density along the symmetry axis for different values of $\alpha_{ano}$. The increase in anomalous transport results in a lower, faster-decreasing electron density in the plume. When comparing $\alpha_{ano}=0.015$ to $\alpha_{ano}=0.018$, a density difference is observed approaching a factor 2 in the plume.
The effects of $\nu_q$ on the plasma density are less noticeable.
%
%

Figure \ref{fig:sensitivity}(b) displays the effect of $\nu_q$ on $T_e$ along the axis. 
As $\nu_q$ increases, the electron cooling rate in the parallel direction increases, and the peak of $T_e$ near the ECR surface in the MN becomes more pronounced.
Electron cooling in a MN without downstream power deposition is well known and has been studied using kinetic models \cite{mart11,mart15a,sanc18b,meri21a}; in this case, power deposition near the ECR surface is noticeable and affects the results of the simulations, as indicated above.
The qualitative effect of increasing $\alpha_{ano}$ on the electron temperature is a nearly uniform upward displacement of the curve driven by changes in the plasma density, and hence the EM power deposition per electron.

In summary, it is found that the simple phenomenological models used here, which rely on uniform values of $\alpha_{ano}$ and $\nu_q$ in the full simulation domain, perform reasonably well in the quest to reproduce the trends of experimental results. However, the absolute values do not match the measurements exactly (although they generally fall within the measurement error). It is worth mentioning that more complex turbulent models (e.g. with nonuniform maps) could be used, but at the probable cost of overfitting and loss of generality across similar devices and different operating points.

\section{Detailed analysis of the HPT discharge}\label{sec:analysis}

Once the simulation parameters have been fitted to agree reasonably well with the limited experimental results, we analyze in detail the plasma discharge
for the case
$\dot{m} = 12.5$ Xe sscm $\equiv 1.23$ mg/s.
The characteristic values / ranges of the plasma conditions found in the simulation domain are shown in Table \ref{table:simparams}.

\subsection{Wave-Plasma Interaction and Electromagnetic Power Deposition}\label{sec:wave}
   
\begin{figure}[ht!]
    \centering
    \includegraphics[width=1\textwidth]{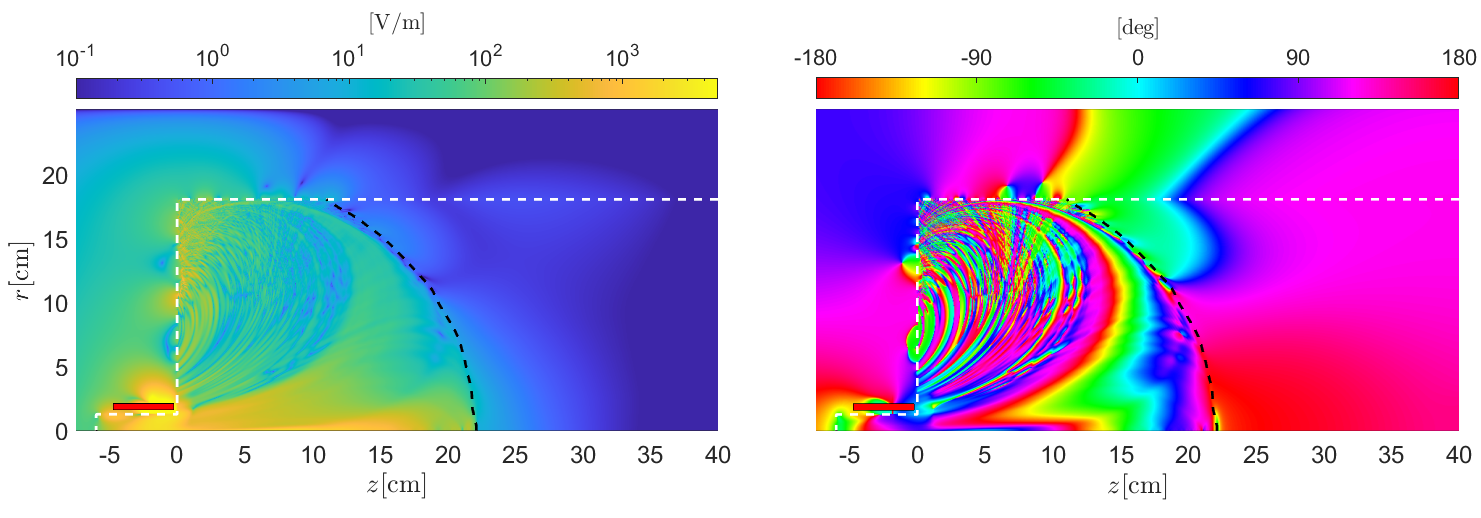}
    \caption{$E_{\theta}^1$ field magnitude (left) and phase angle (right). The dashed black line corresponds to the ECR surface.}
    \label{fig:wavefields}
\end{figure}

The magnitude and phase of the $m=1$ azimuthal electric field of the W-module solution, which is the predominant component excited by helical antennas \cite{chen97a}, are shown in Figure \ref{fig:wavefields}. 
The field is strongest inside the source, decaying in the plume and thruster surroundings. 
Despite the lower magnitude, propagation takes place in the MN up to the ECR surface. Beyond it, since the plasma is overdense for the applied frequency, the fields become evanescent \cite{STIX92,jime22a}. 

Along the axis, a low $k_\parallel$ helicon mode is seen to propagate in the source and near-plume.
The plume periphery presents shorter wavelengths, consistent with the Trivelpiece-Gould waves expected at the plasma edge \cite{blac02}. These structures exhibit a high perpendicular wave number $k_\perp$. 
Both kinds of wave (helicon and Trivelpiece-Gould) correspond to the right-hand polarized whistler wave at different propagation angles with respect to $\bm B$, and both vanish upon arriving at the ECR surface. At this location, electromagnetic power is mainly absorbed and cannot propagate beyond. On a side note, we mention that there is another ECR surface in this simulation setup in the neighborhood of the singular point inside the plasma vessel. However, its effects are minor due to the limited spatial extent of that transition.

\begin{figure}[ht]
\centering
\begin{minipage}[c]{0.49\textwidth}
\includegraphics[width=0.95\textwidth]{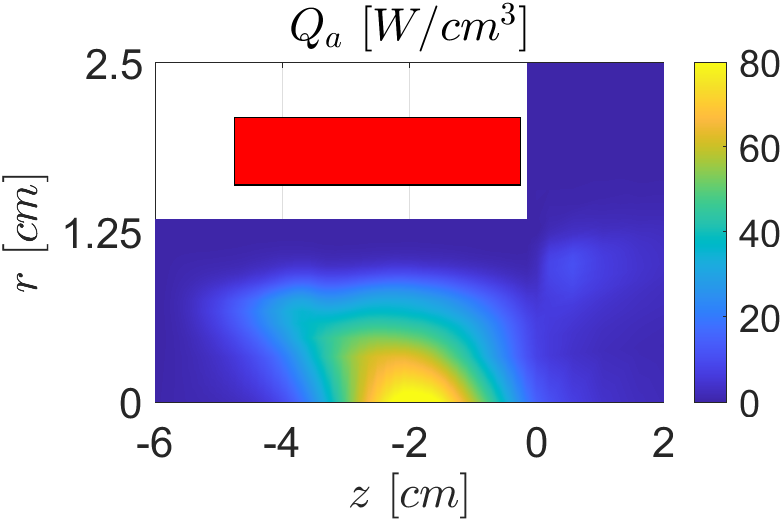}
\end{minipage}
\begin{minipage}[c]{0.49\textwidth}
\includegraphics[width=1\textwidth]{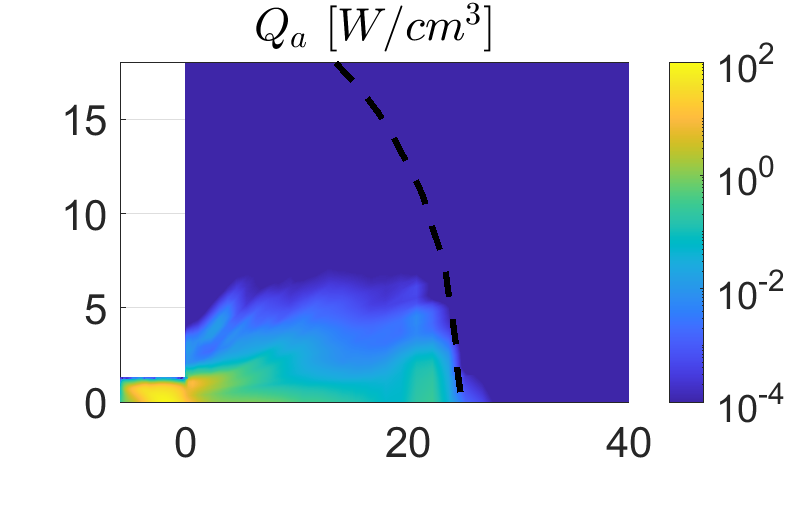}
\end{minipage}

\begin{minipage}[c]{0.49\textwidth}
\includegraphics[width=1.02\textwidth]{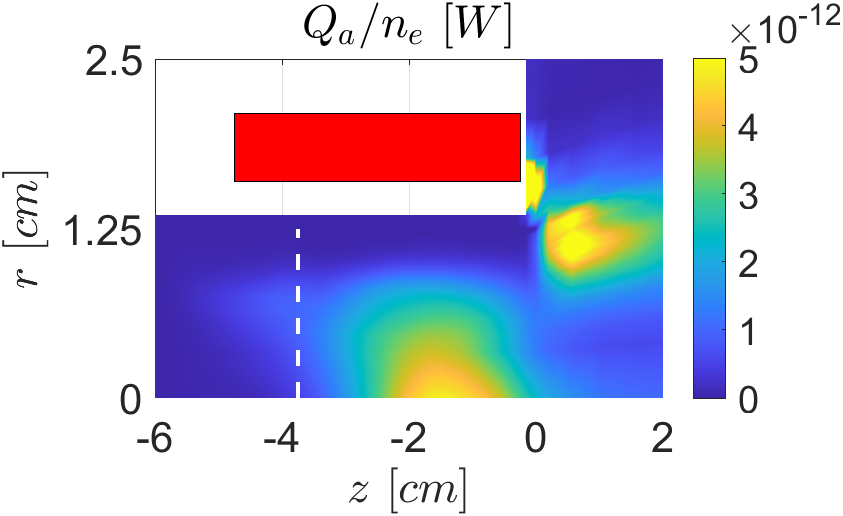}
\end{minipage}
\begin{minipage}[c]{0.49\textwidth}
\hspace{2pt}
\includegraphics[width=1\textwidth]{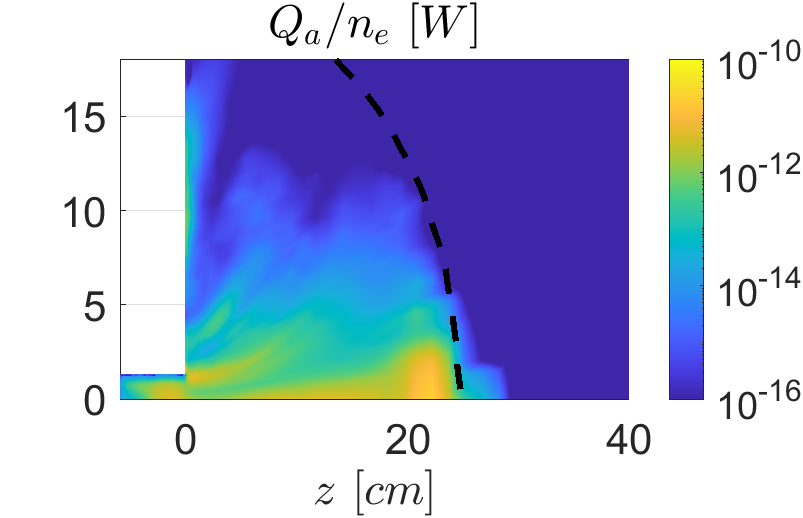}
\end{minipage}

\caption{Electromagnetic Power Deposition (top row), power per electron (bottom). Zoom on the source region (left column) and full transport domain (right column). The magnetic field separatrix is shown as a white dashed line in the zoom view plots. The ECR resonance is shown as a black dashed line in the full view plots.
}
\label{fig:power}
\end{figure}

The power absorption profile is shown in the top row of Figure \ref{fig:power}. Due to the linearity of the wave problem, these maps can also be interpreted as the local plasma resistivity (save for a factor). Most of the power is absorbed within the source. This agrees well with plasma-wave theory, predicting an absorption approximately proportional to the plasma density for similar magnitude propagating fields
and suggests a good coupling between the RF antenna and the plasma inside the source.
In the zoomed-in view of the source, we see that the maximum deposition is located near the axis, peaking at the vicinity of the magnetic separatrix.  

The bottom row of Figure \ref{fig:power} shows the power absorbed \textit{per electron}. 
According to the electron energy equation, this quantity determines the local heating of electrons, and thus affects to a large extent the profile of $T_e$. Per-particle absorption peaks inside the source but spreads into the plume, with local maxima at the vicinity of the ECR surface near the symmetry axis, revealing the importance of the particle-specific power deposition in that region of the MN. There is no energy deposition downstream of the ECR surface.
Although earlier work has relied on computing absorption only within the thruster plasma source\cite{souh21}, these results suggest that local heating may affect the structure of the plume and drive the electron temperature profile outside the source.

\begin{figure}[ht]
    \centering
    \includegraphics[width=0.45\textwidth]{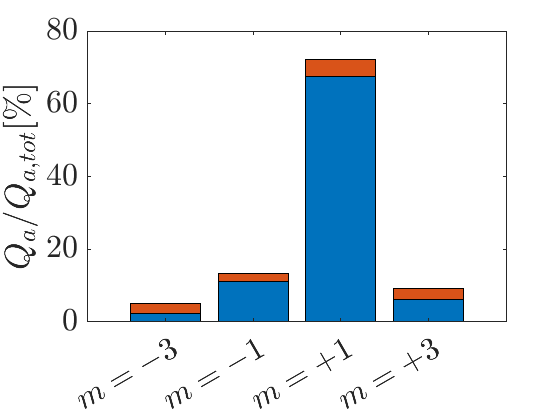}
    \caption{Fraction of the total electromagnetic power deposition in the source (blue) and plume (orange) regions for each azimuthal wave number.}
    \label{fig:modes}
\end{figure}

So far, we have exclusively focused on the fields carried by EM waves in azimuthal mode $m=1$. This was done on the basis of extensive past evidence on the predominant power coupling of helical antennas
in helicon sources \cite{arnu98}. However, wave behavior is known to be highly dependent on the applied magnetic field \cite{STIX92}, and the question arises as to the extent to which the previous mechanisms of propagation and absorption would still be valid in our cusped-field topology. Figure \ref{fig:modes} shows the fraction of the total power deposition contributed by each of the four modes considered in the steady state discharge.  Although the power deposited by $m=-1$ is higher than in a classical topology \cite{jime22a}, it can be seen that $m=1$ is still the main contributor.
As suggested by the results for $m=\pm3$, the power of the higher modes decreases rapidly with $m$. 
Additionally, the majority of the power is deposited in the dense plasma of the source regardless of the mode. 
Therefore, we conclude that our approximation, that is, considering only $m=1$ in coupled simulations, is justified, with a loss of accuracy comparable to other problem uncertainties such as wave and transport coupling or antenna modeling.

\subsection{2D Plasma Discharge Profiles}\label{sec:prof}

\begin{figure}[p]
\centering
\vspace{-1.0cm}
\begin{minipage}[c]{0.48\textwidth}
\includegraphics[width=1\textwidth]{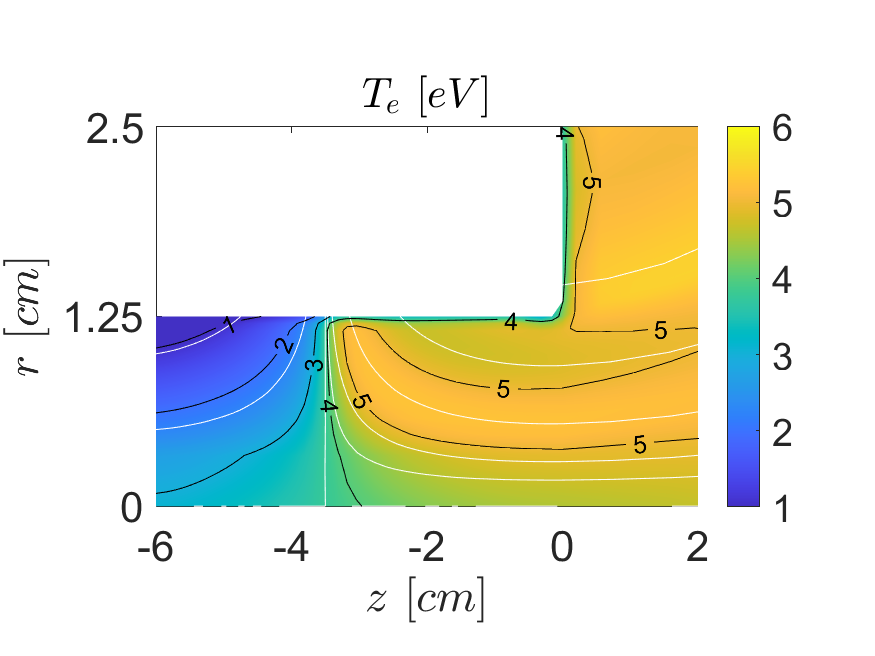}
\vspace{-2.00cm}\hspace{2 cm}
\end{minipage}
\begin{minipage}[c]{0.48\textwidth}
\includegraphics[width=1\textwidth]{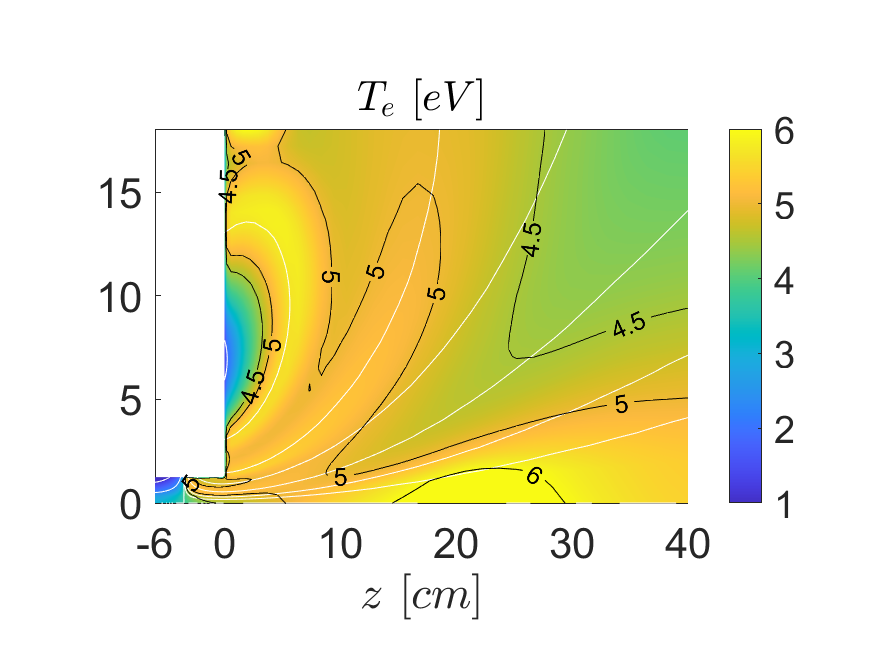}
\vspace{-2.00cm}\hspace{2 cm}
\end{minipage}
\begin{minipage}[c]{0.48\textwidth}
\includegraphics[width=1\textwidth]{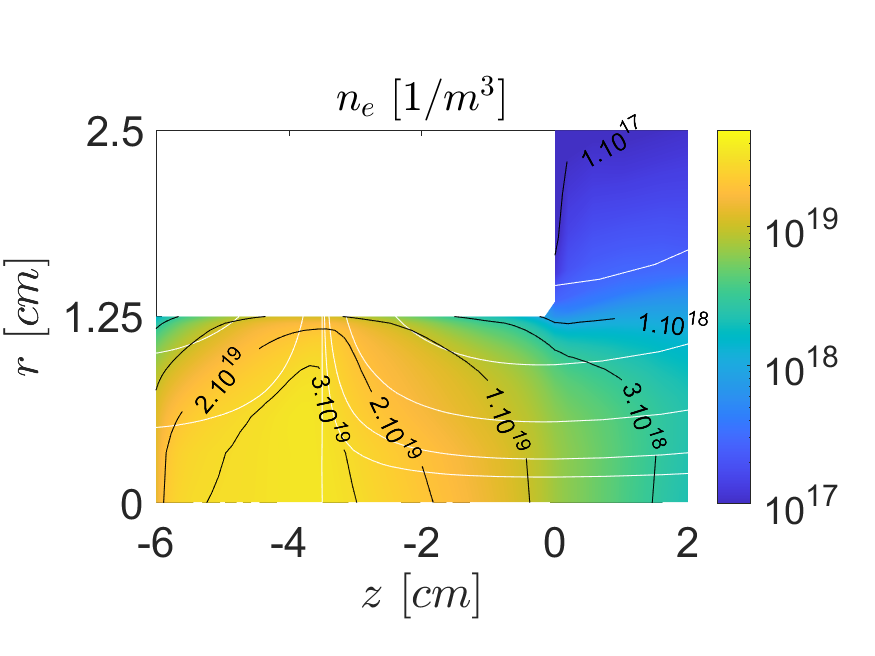}
\vspace{-2.00cm}\hspace{2 cm}
\end{minipage}
\begin{minipage}[c]{0.48\textwidth}
\includegraphics[width=1\textwidth]{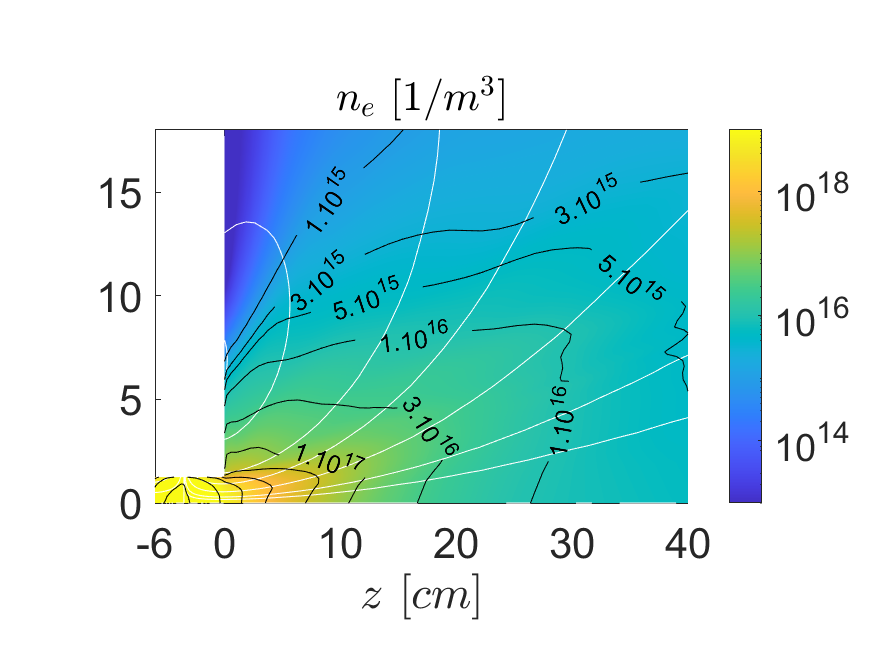}
\vspace{-2.00cm}\hspace{2 cm}
\end{minipage}
\begin{minipage}[c]{0.48\textwidth}
\includegraphics[width=1\textwidth]{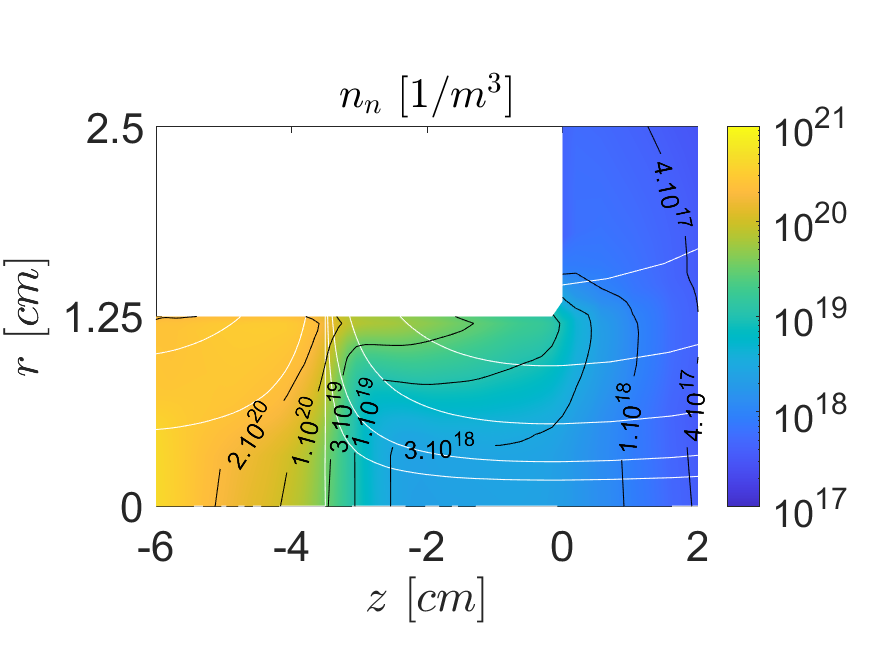}
\vspace{-2.00cm}\hspace{2 cm}
\end{minipage}
\begin{minipage}[c]{0.48\textwidth}
\includegraphics[width=1\textwidth]{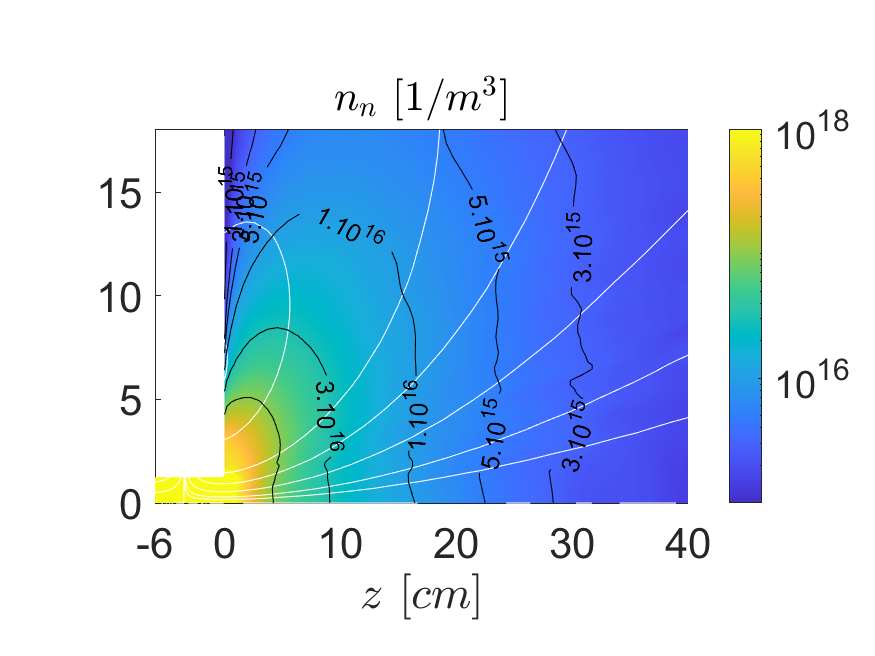}
\vspace{-2.00cm}\hspace{2 cm}
\end{minipage}
\begin{minipage}[c]{0.48\textwidth}
\includegraphics[width=1\textwidth]{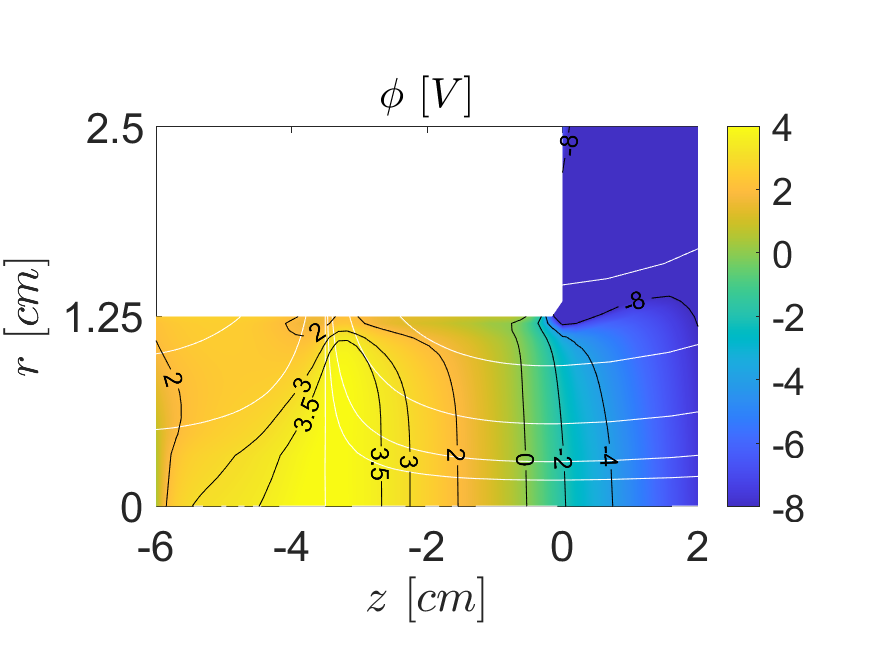}
\vspace{-0.25cm}\hspace{2 cm}
\end{minipage}
\begin{minipage}[c]{0.48\textwidth}
\includegraphics[width=1\textwidth]{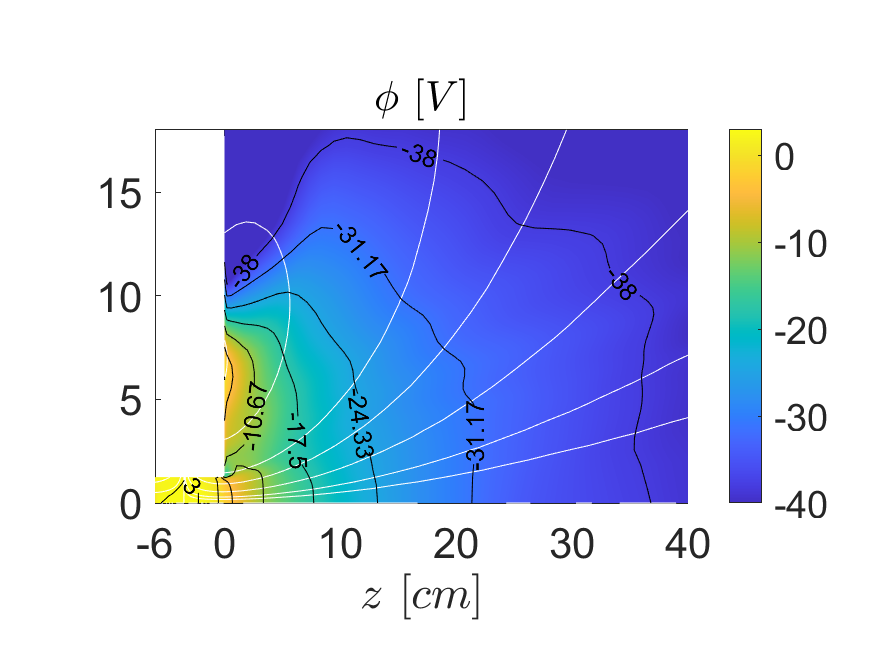}
\vspace{-0.25cm}\hspace{2 cm}
\end{minipage}

\caption{Electron temperature (1st row), plasma density (2nd row),  neutral density (3rd row), and plasma potential (4th row). Zoom in on the source region (left column) and full transport domain view (right column). The solid white lines are magnetic field streamlines.}
\label{fig:prop}
\end{figure}

Figure \ref{fig:prop} shows 2D maps of the main magnitudes of the plasma response.  Some of the principal features are qualitatively similar to those reported in \cite{zhou22a} for an EPT with a quasi-axial magnetic field and localized power deposition inside the thruster. 
Here, the cusped magnetic topology and the more spread power deposition give rise to some peculiarities. 

The electron temperature $T_e$ inside the source in Figure \ref{fig:prop}
responds mainly to the power deposition map, which, as shown in Figure \ref{fig:power}, is centered at the antenna location and mainly downstream of the magnetic cusp.
The radial magnetic lines near the cusp separate hot and cool very efficiently,
resulting in a stark difference in $T_e$ left and right of the cusp: while $T_e$ after the cusp is around $5$ eV, it is below $2$ eV left of it.
Naturally, $T_e$ is also affected by wall losses, inelastic collisions, and convection/heat fluxes.
As parallel transport is very efficient, we observe a near-isothermal behavior along magnetic lines on this scale. 

The low $T_e$ inwards of the magnetic cusp
means low plasma production and neutral depletion there, as the plot of $n_n$ in Figure \ref{fig:prop} suggests. Ionization of neutrals becomes efficient as $T_e$ increases when crossing the cusp and neutrals are depleted.
Additionally, the neutral density rises locally near the walls, driven by ion surface recombination.
The plasma density peaks around the magnetic cusp section, but unfortunately, this plasma is not transported exclusively downstream, but upstream as well, which is found to be a major loss mechanism of this HPT field configuration.

\begin{figure}[ht]
    \centering
    \includegraphics[width=\textwidth]{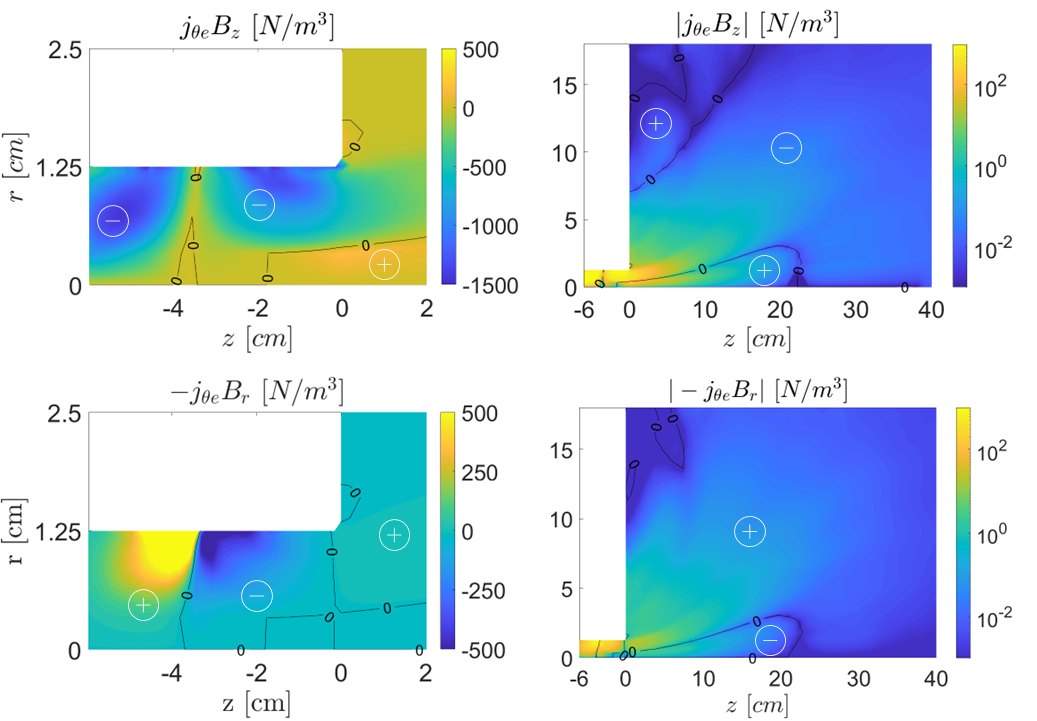}
    \caption{Radial (top) and axial (bottom) magnetic force in the plasma. The force direction is indicated by $\pm$ signs.}
    \label{fig:fm}
\end{figure}

In Figure \ref{fig:prop}, the map of the electrostatic potential $\phi$ 
inside the thruster roughly follows that of $\ln n_e$.
However, the peak of $\phi$ occurs downstream from the peak of $n_e$.
The large plasma production near the peak of $\phi$ flows both downstream and upstream, 
facilitated by the axial magnetic lines guiding electron motion and the unmagnetized character of the ions. 
Downstream the cusp, the decrease of $n_e$ is due to plasma acceleration.
Except in the proximity of the lateral wall, the electric field $-\nabla \phi$ becomes essentially axial where the magnetic lines are axial. The low radial electric field is a consequence of the existence of an azimuthal electron current that produces a radial magnetic force, shown in Figure \ref{fig:fm} that compensates for pressure gradients \cite{ahed09k, ahed13c, zhou22a}.

The map of $T_e$ in the plume region, shown in the top left of Figure \ref{fig:prop},  also merits discussion.
The electron temperature remains around its maximum value in the near-plume due to the mild cooling rate.
Although the gradient of $T_e$ is also mainly in the perpendicular direction in the plume, certain parallel variations can be observed at this scale.
The local maximum near the ECR location on the axis, reported in Section \ref{sec:fitting}, is consistent with the peak in power deposition per electron there, which helps raise the local $T_e$.
Another local maximum of $T_e$ occurs in the periphery of the plume, on magnetic lines that do not connect to the plasma source. This behavior, which has not yet been validated experimentally, is driven by low $n_e$, low energy losses, and low perpendicular transport in this region, combined with moderate RF power deposition per particle. 
Finally, the electrostatic potential map in the plume of Figure \ref{fig:prop} also features a prominently axial gradient, as inside the source.
This behavior suggests a low plume divergence, as discussed in the next section.

\begin{figure}[ht]
    \centering
    \includegraphics[width=0.5\textwidth]{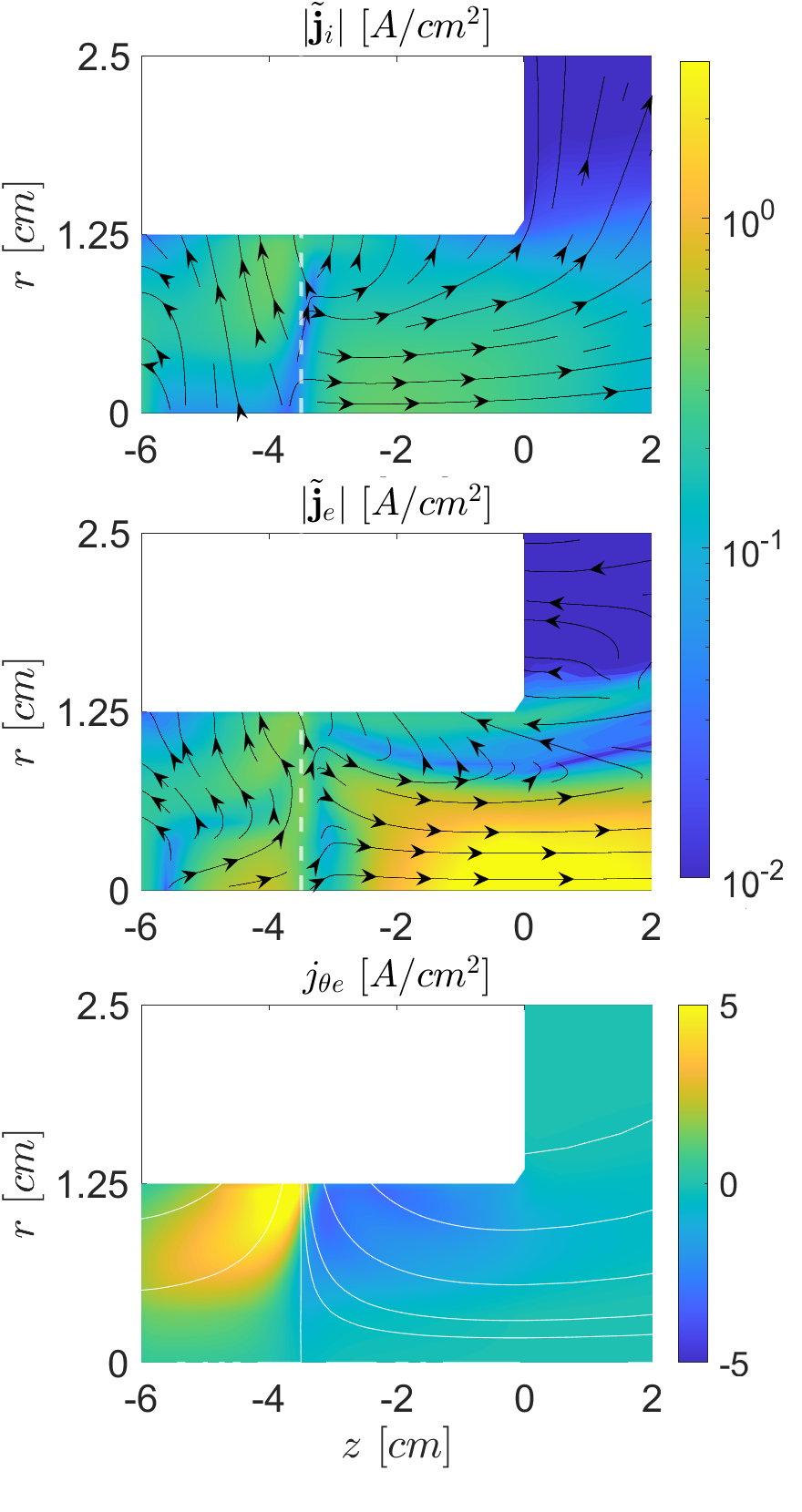}
    \caption{Total ion (top) and electron (middle) current density and particle flux vectors; and azimuthal electron current (bottom). The magnetic field separatrix is shown as a white dashed line.}
    \label{fig:currents}
\end{figure}

Figure \ref{fig:currents} shows the in-plane ion flux (top, dominated by singly charged ions) and the total electron flux (middle), $\boldsymbol{\tilde \jmath}_i = j_{zi}\bm 1_z + j_{ri}\bm 1_r$ and $\boldsymbol{\tilde \jmath}_e = j_{ze}\bm 1_z + j_{re}\bm 1_r$.
There is a significant flux of particles into the back and lateral walls of the source.
Moreover, very few ions generated upstream of the magnetic cusp separatrix are capable of crossing it. As a consequence, ion production to the left of the cusp is almost completely lost to the dielectric walls.  Considering that ions are only weakly magnetized, the ion dynamics is driven solely by the potential map.
Downstream of the separatrix, ions accelerate and expand, with a nonnegligible fraction of them impacting the lateral wall. 
Electrons, on the other hand, feature a large axial current near the axis and a compensating return current at larger radii to keep the plasma emission globally current-free.

The azimuthal current densities are such that $j_{\theta e}\gg j_{\theta i}$ by about a factor 200, and the electron contribution constitutes the bulk of the plasma azimuthal current, essential in magnetic confinement and thrust generation. 
The map of $j_{\theta e}$ at the source is shown in
Fig. \ref{fig:currents} (bottom).
From the momentum equation \eqref{eq:moment}, 
the perpendicular pressure gradient, the perpendicular electric field, and to a minor extent, classical resistivity contribute to determining this current. As can be observed, $j_{\theta e}$ peaks near the lateral wall around the location of the cusp, switching signs to each side of it.

The radial and axial magnetic force densities generated by the interaction of this azimuthal current with the applied magnetic field, respectively $j_{\theta e} B_z$ and $-j_{\theta e}B_r$, are shown in Figure \ref{fig:fm}. 
The radial magnetic force is negative in the source and confines the electrons away from the lateral wall, whereas the axial magnetic force is positive and negative left and right of the cusp, respectively. 
Combined, the two forces push the electrons to the singular point of the field, where the highest plasma density $n_e$ occurs.

From the point of view of magnetic thrust generation (that is, the volume integral of $-j_{\theta e}B_r$), it is evident that the electron currents to the left of the cusp contribute positively to thrust, while those to the right of the cusp exert a comparable negative (drag) contribution, driven by the geometry of the field and the direction of the electron pressure and electrostatic potential gradients. 
The axial magnetic force in each of these two regions amounts to approximately $4.7$ and $-2.9$ mN, respectively.
The positive thrust generation is resumed downstream in the MN, where $3.2$ mN are generated up to the end of the simulation domain. This is roughly $45\%$ of the total thrust of the device, a fraction comparable to that of other MNs \cite{meri16g}. In addition to the magnetic thrust, the pressure thrust (on the wall of the plasma source) yields about $2.2$ mN extra.

\begin{figure}[ht]
\centering
\begin{minipage}[c]{0.45\textwidth}
\hspace{-0.3cm}
\includegraphics[width=\textwidth]{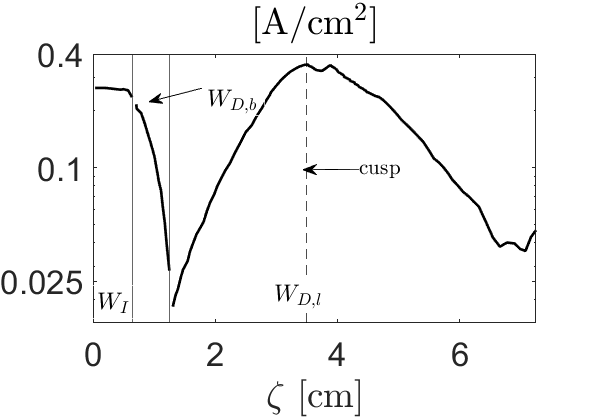}
\end{minipage}
\begin{minipage}[c]{0.45\textwidth}
\includegraphics[width=0.97\textwidth]{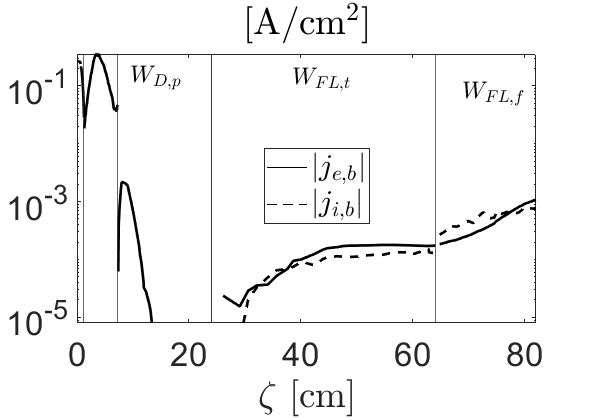}
\end{minipage}
\begin{minipage}[c]{0.45\textwidth}
\vspace{0.2cm}
\includegraphics[width=0.95\textwidth]{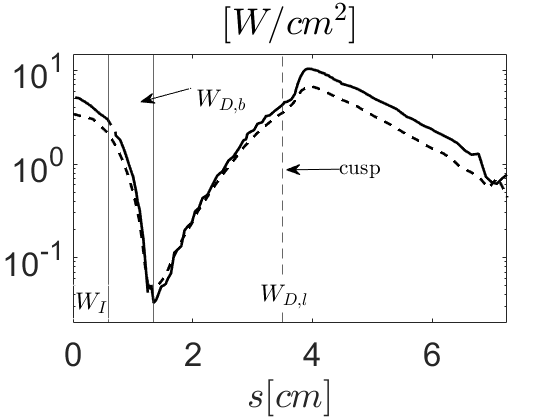}
\end{minipage}
\begin{minipage}[c]{0.45\textwidth}
\vspace{0.2cm}
\includegraphics[width=0.95\textwidth]{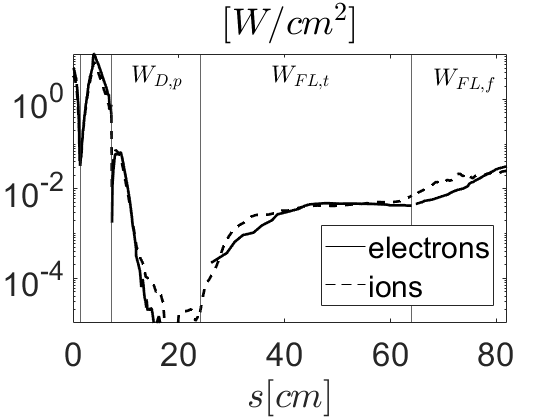}
\end{minipage}
\caption{
Electron and total ion current densities (top row) and energy fluxes (bottom row) to source dielectric walls (left column) and all domain boundaries (right column). The arc length coordinate $s$ moves clockwise along the boundary from $(z,r)=(-6,0)$.  Refer to Figure \ref{fig:sim_setup} to identify each boundary wall $W$. The dashed line in the left plots indicates the location of cusp.}
\label{fig:bcurrents}
\end{figure}

Figure \ref{fig:bcurrents} shows the electron and total ion fluxes to the boundaries of the domain. While these fluxes differ at the top and right plume boundaries, their integral satisfies the global current-free condition. 
Figure \ref{fig:bcurrents} (right) shows that the maximum current is reached just at $z = -3.5$ cm, demonstrating not only the poor confinement in the source, but also that the cusped magnetic field drives the particles directly onto the walls at that location.

The energy fluxes in the bottom row of Figure \ref{fig:bcurrents} follow similar trends. In most of the inner wall of the plasma source, the energy carried by the electrons (advective and heat flux terms) is larger than the ion energy. Poor confinement in the neighborhood of the magnetic separatrix drives a strong convective energy flux to the lateral wall. This important energy sink and the need to reionize the particles that recombine at the walls lead to the low plasma temperature shown in Figure \ref{fig:prop}. Secondary electron emission is small, with a yield less than 20 $\%$ in all source walls, due to the moderate electron temperature and ion impact energies.

\FloatBarrier

\subsection{Balances of plasma production and energy}\label{sec:perf}

The efficiency of the thruster is analyzed here through different balances. The first variable of interest is
the total ion mass production rate, $\dot m_{i,prod}/\dot m$. 
In steady state, this ion production is distributed 
as 
\begin{equation}
    \dot m_{i,prod}=\dot m_{i\infty}+\dot m_{i, lat}+\dot m_{i, back}
    \label{eq:massbalance}
\end{equation}
with $\dot m_{i\infty}$ the flow of ions through the free-flow boundaries and $\dot m_{i, lat}$ and $\dot m_{i, back}$ the ion flows that recombine at the lateral and back wall, respectively.
For the case simulated here, the utilization of the propellant mass is $\eta_u=\dot m_{i\infty}/\dot m\simeq 89\%$
and the production efficiency is $\eta_{prod}=\dot m_{i\infty}/\dot m_{i, prod}\simeq 9\%$; both values are coherent with the high densities and the large chamber length.
This means that plasma ionization is very efficient, but plasma confinement is very poor. Indeed, on average, every neutral is ionized 10 times after 9 recombinations in the walls.
A balance equivalent to \eqref{eq:massbalance} can be done in terms of total ion electric currents, that is,
$I_{i,prod}=I_{i\infty}+I_{i, lat}+I_{i, back}$, which results from integrals of the current in Fig. \ref{fig:bcurrents}.
Ratios for mass flows and currents are very similar 
since the amount of doubly-charged ions is small; for instance, the downstream ion mass flow consists of 91.5\% of singly-charged ions and 8.5\% of doubly-charged ions.

The power balance can be expressed as 
\begin{equation}
    P_a=P_\infty+P_{lat}+P_{back}+P_{ion}+P_{exc}
\end{equation}
with the terms on the right hand side being, respectively, the mechanical power in the beam leaving the plume domain, the mechanical power lost in the lateral walls and the back walls of the source, and the power spent in ionization and excitation collisions. The first three are integrals along sections of the curves of Figure \ref{fig:bcurrents}.
The ratios to the absorbed power $P_a$ are the following:
\begin{equation}
   \frac{P_\infty}{P_a}\simeq 12\%,
   \quad
   \frac{P_{lat}}{P_a}\simeq 33\%,
   \quad
   \frac{P_{back}}{P_a}\simeq 1.6\%,
   \quad
   \frac{P_{ion}}{P_a}\simeq 22\%,
   \quad
   \frac{P_{exc}}{P_a}\simeq 31\%,
\end{equation}
%
i.e. a 35\% of $P_a$  is lost in the walls and a 53\% is lost in ionization and excitation. These large losses are due to the combination of a poor plasma confinement and a very long chamber. Despite the relatively low $T_e$ (about 5 eV), there is multiple ion recombination and neutral reionization, as illustrated by the production efficiency of only 9\%. The low fraction of useful energy explains the low $T_e$, which leads to excitation energy losses being higher than ionization ones.

It is of interest to compare these performances with those of
Ref. \cite{zhou22a} for an HPT of similar geometry and power, but with a quasi-axial magnetic topology. 
The ratio of back to lateral wall areas is about 10\% in the two cases. The ratio of back to lateral wall energy losses is about 4.5\% here and it was about 50\% in Ref. \cite{zhou22a}. Thus, the magnetic cusp has improved the
confinement of the back wall by moving plasma production downstream of the ring cusp. However, the overall confinement is worse, as illustrated by $\eta_{prod}$, which is 16\% (almost double) in \cite{zhou22a}.
In that case,  44\% of $P_a$  is lost in the walls and a 36\% is lost in ionization and excitation. The higher percentage of wall losses in \cite{zhou22a} is due to the higher $T_e$ (about 11 eV) that also comes with a considerable reduction in excitation losses. This is again an indication of the complex energy balance in these thrusters and the
delicate optimization of design and operation conditions.

The thrust calculated here is $F\simeq 7.2$ mN, which yields a thrust
efficiency $\eta_F=F^2/(2 \dot mP_a)\simeq 5.1\%$, 
(compared to 11.5\% in \cite{zhou22a}).
To understand the different sources of thrust inefficiency, $\eta_F$ is factorized as \cite{zhou22a}:
\begin{equation}
\eta_F=\eta_{disp}\eta_{div}\eta_{ene},
\quad
\eta_{disp}=\frac{F^2}{2\dot m P^{(z)}_\infty},
 \quad
 \eta_{div}=\frac{P^{(z)}_\infty}{P_\infty},
\quad
\eta_{ene}=\frac{P_\infty}{P_a},
\end{equation}
where: $P^{(z)}_\infty$ is the flow of axial energy in the beam, the only one that may be used to generate thrust; $\eta_{disp}$ takes into account that the plasma beam is not monoenergetic, and constituted of particles of different electric charges; $\eta_{div}$ assesses the plume divergence; and $\eta_{ene}$ is the energy efficiency.
For our simulation we have:  
$\eta_{ene}\simeq 12\%$, as mentioned above, which is the main cause of the poor efficiency of the device;
$\eta_{div}\simeq 59\%$, which corresponds to a divergence half angle of $\arccos\sqrt{\eta_{div}}\simeq $ 40\ deg; and
$\eta_{disp}\simeq 72\%$, value that comes out mainly from 
the non-monoenergetic downstream plasma flow, with an  11\% 
constituted of slow neutrals
and a 7\% of doubly-charged ions, 
with double energy than the 82\% of 
singly-charged ions.

\section{Conclusion}\label{sec:concl}

The discharge of a cusped Helicon Plasma Thruster
has been analyzed through experimental and simulation results. 
First, experimental plasma profiles in the plume have been reported and used to tune some anomalous collisionality parameters of the 2D simulation code HYPHEN-EPT.

The simulations are capable of capturing the trend of the main plasma variables, including the presence of a local maximum of electron temperature $T_e$ near the downstream ECR surface, although differences of up to 50\% in absolute value are observed in the high mass flow rate case. 
The ion current density $j_i$ is also well described by the code up to moderate angles, beyond which the experimental data present lateral wings, possibly due to facility effects, not included in the simulation.
The results show a high sensitivity to the parameters that define electron anomalous cross-field transport and far-plume cooling.

The numerical model provides ample information on the different sources of energy and performance losses. Plasma currents to walls are very large, implying high ion recombination, multiple re-ionization, which in turn increase the losses due to ionization and excitation. This results in low energy efficiency and, consequently, low thrust efficiency. 
One goal of this study has been to compare this cusped magnetic topology with the conventional, quasi-axial one. The presence of a ring-cusp can improve the wall losses in certain regions of the source, but the overall effect on thrust is negative, at least for the cases considered here. Hence, increasing the thrust efficiency of the HPT requires mainly further investigation of geometric and magnetic designs to better confine the plasma.

The magnetic ring cusp plus the location of the antenna determines much the plasma response inside the source. 
In this HPT configuration, the deposition of energy is located close to the center of the antenna, which is downstream of the cusp.
This causes the region inward to the cusp to be of low temperature and displaces to the cusp the main ionization region. 
The cusp also divides the regions of back and forward ion currents, and of positive (inward) and negative (outward) magnetic thrust inside the source; the external magnetic thrust is positive again in the MN. 
Overall, an encouraging result is that the net magnetic thrust amounts to 70\% of the total generated thrust, and the conclusion is that a parametric study on the relative locations of the antenna and cusp is also fundamental for optimizing the HPT.

Finally, a numerical contribution has been the improvement of the W-module of HYPHEN-EPT with new Finite Element algorithms capable of reproducing any azimuthal wavenumber $m$. These are shown to be more accurate than the previous Finite Difference ones. For the half-helical antenna of the HPT, $m=+1$ is the dominant mode (with $\sim$70\% of the deposited power), and the modes $m=-1$ and $+3$ are subdominant. The computed wavefields indicate that most of the RF power is absorbed inside the source and downstream of the magnetic cusp. Furthermore, the power absorbed per electron is seen to have a local maximum on the axis near the downstream ECR surface, which correlates with the peak in $T_e$ observed in experiments and simulations. Being relatively far from the plasma source, the role of ECR in plasma heating was ignored in previous HPT works (e.g. \cite{zhou19b, souh21}).

\section*{Acknowledgments}
 
The authors would like to thank Adrian Dominguez-Vazquez for his useful comments, especially on the use of the HYPHEN simulation code. 
Initial support for the activities leading to work came from the HIPATIA project, funded by the European Union’s Horizon 2020 Research and Innovation Program, under Grant Agreement number 870542. 
The completion of this work was carried out with funding from the European Research Council (ERC) under the European Union’s Horizon 2020 research and innovation programme (grant agreement No 950466, project ZARATHUSTRA).

\appendix

\section{Finite element vs finite differences discretization for the W-module}\label{sec:appendix}

\begin{figure}[ht]
\centering
\includegraphics[width=0.8\textwidth]{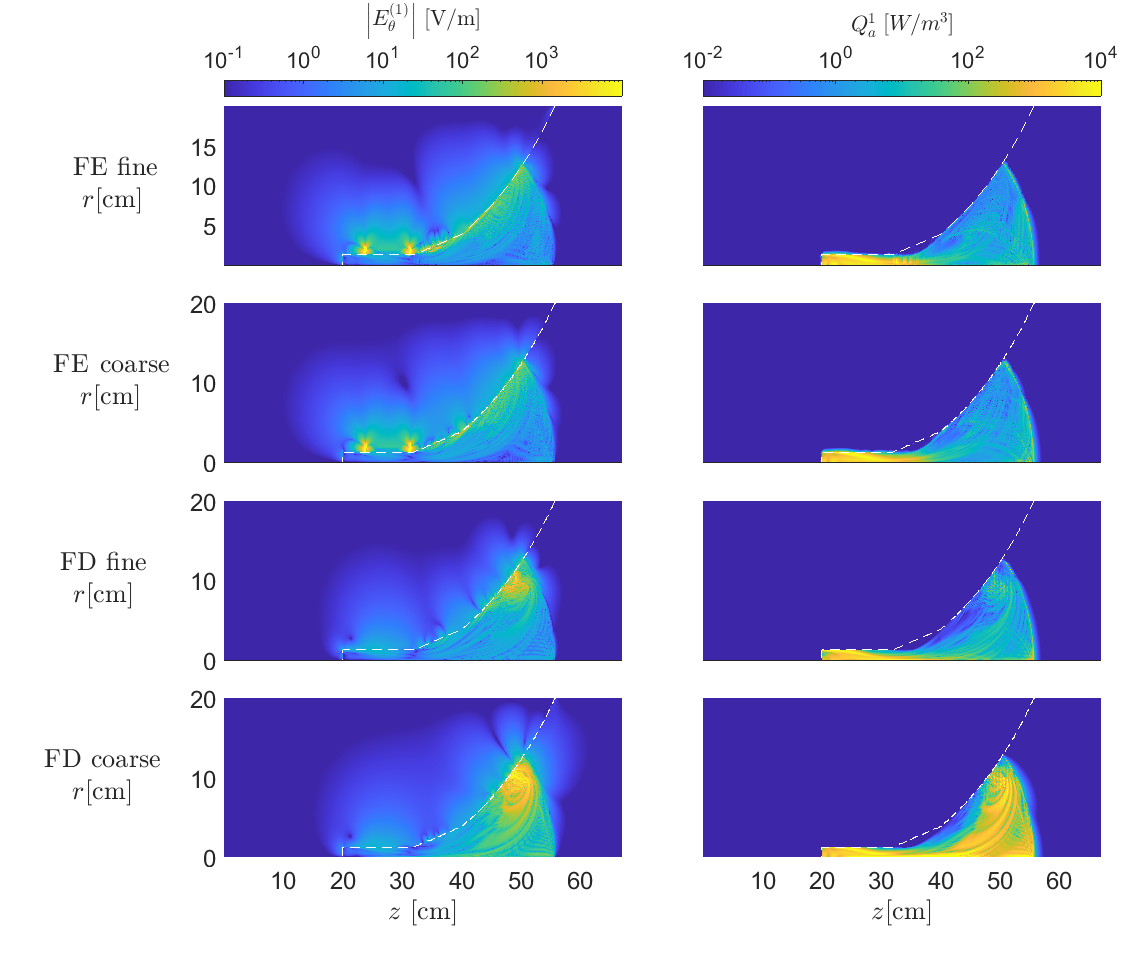}
\caption{Finite Element (FE) and Finite Difference (FD) full thruster and plume wave solutions \cite{jime22a}. $E_{\theta}^1$ electric field magnitude (1st column), and power deposition (2nd column). Putative spurious noise is localized near the $\omega=\omega_{pe}=\omega_{ce}$ crossing in the FD simulations and, in the coarse case, extends throughout most of the plume region. Coarse meshes have about $\sim 3\cdot10^5$ nodes and fine meshes $\sim 1.4\cdot 10^6$ nodes.
}\label{fig:FEMvsFD}
\end{figure}

Previous full-wave solvers \cite{chen12b,tian18a} applied to Helicon thrusters have relied on Finite Difference (FD) discretizations of Maxwell's equations, predominantly based on structured staggered grids and Yee's method \cite{yee66}.

The presence of a full dielectric tensor for waves traveling through a magnetized plasma medium introduces the need for interpolations in the electric field components placed at different nodes of the staggered grid. These interpolations can induce spurious noise in transition regions where the dielectric tensor does not vary smoothly \cite{jime22a}. Additionally, the quality of the solution seems to depend on the relative alignment of the mesh and the applied magnetic field,
and, moreover, the spatial variation of the dielectric tensor with plasma properties calls for widely different mesh sizes at locations with long and short wavelengths.

A conforming Finite Element (FE) discretization on an unstructured grid
allows us to partially tackle these problems.
Following the work of Sanchez et al. \cite{svil21b}, in this work Nédeléc vector elements are used for the in-plane fields and Lagrange elements for the out-of-plane field.
The new implementation is extended to arbitrary azimuthal $m$ modes and is based on the FEniCSx library \cite{fenics, scro22} rather than MFEM \cite{mfem} as in \cite{svil21a}, motivated by the support of complex numbers. 

The performance of the new FE code is compared with the older FD code from \cite{jime22a}, for the same simulation setup of that work. 
Figure \ref{fig:FEMvsFD} shows the magnitude of the azimuthal field and the power deposition map for $m=1$. 
Although the FE results are essentially invariant under two different mesh sizes and therefore we can claim mesh convergence, this is not true for the FD results with comparable or even finer grids. Furthermore, the FD results feature spurious large field values near the triple point where $\omega=\omega_{pe}=\omega_{ce}$, around $z=50$ cm, $r = 10$ cm. 
This numerical error is prevalent in most of the plume region and clearly affects the power deposition maps.

In addition to the increase in accuracy and the elimination of spurious features, the efficiency of the FE solver is much higher than that of the FD one, achieving an average reduction factor of 5 in the CPU wall time with comparable meshes. These advances come mainly from the parallelization of the FE library and the solvers used.  
 
\bibliography{ep2, others}

\end{document}